\documentclass[12pt]{article}
\usepackage{epsfig}
\usepackage{hyperref}
\usepackage{graphicx}
\usepackage{ragged2e}
\usepackage{cite} 
\usepackage{slashed}

\newcommand{\be}{\begin{equation}}
\newcommand{\ee}{\end{equation}}
\newcommand{\bea}{\begin{eqnarray}}
\newcommand{\eea}{\end{eqnarray}}

\newcommand{\Sig}{{\bf\Sigma}}

 \newcommand{\Sigb}{{\bf{\overline\Sigma}}}

\newcommand{\nnu}{\nonumber\\}
\def\blfootnote{\xdef\@thefnmark{}\@footnotetext}

\begin{document}
\vspace{4\baselineskip}
\vspace{4cm}
\begin{center}{\Large\bf Grand Pleromal  Transmutation :  
 UV Condensates via Konishi Anomaly, Dimensional
Transmutation and Ultraminimal GUTs}

\end{center}
\vspace{2cm}
\begin{center}
{\large
Charanjit S. Aulakh\footnote{aulakh@iisermohali.ac.in} 
 }
\end{center}
\vspace{0.2cm}
\begin{center}
 Indian Institute of Science Education and Research
Mohali,\\ Sector 81, S. A. S. Nagar, Manauli PO 140306, Punjab India \\
and\\
 International Centre for Theoretical Physics,\\
 Strada Costiera 11, 34100,Trieste,  Italy\footnote{Senior Associate ICTP, 2013-2019}\\
\end{center}

\vspace{.7 true cm}
\centerline{\Large{\bf{Abstract}}}
\vspace{.7 true cm}
{  Using consistency requirements relating chiral condensates  imposed by the so called Generalized Konishi Anomaly,  we show  that  dimensional transmutation via gaugino condensation    {\emph{in the ultraviolet}}   drives gauge symmetry breaking in a  large   class  of   {\emph{asymptotically strong}} Super Yang Mills Higgs  theories. For Adjoint multiplet  type  chiral superfields $\Phi$  (transforming as   $r \times \bar r$ representations of a non Abelian  gauge group G),   solution of the Generalized Konishi Anomaly(GKA)  equations allows calculation of   quantum corrected VEVs in terms of the dimensional transmutation    scale $\Lambda_{UV} \simeq M_X \,  e^{\frac{8\pi^2}{  g^2(M_X) b_0}} $  which determines the   gaugino condensate. Thus  the gauge coupling at the  perturbative unification scale $M_X$   generates    GUT symmetry breaking VEVs  by non-perturbative  dimensional transmutation.   This   obviates  the need for    large(or any)  input   mass scales in the superpotential.    Rank reduction  can be achieved by including   pairs of  chiral superfields  transforming as  either $({\bf Q}(r),{ \bf\bar Q}(\bar r))$   or    $ (\Sig((r\otimes r)_{symm})),  \Sigb(({\bar r \otimes\bar r})_{symm})$, that  form trilinear matrix  gauge invariants $\bar Q\cdot \Phi\cdot Q, \Sigb \cdot  \Phi\cdot \Sig $   with $\Phi$.   Novel, robust and {\emph{ultraminimal}} Grand unification algorithms emerge from the analysis.  We sketch the structure of a realistic Spin(10) model, with the $16$-plet of Spin(10) as   the base representation $r$,   which mimics the  realistic  Minimal Supersymmetric GUT but    contains  even fewer free parameters. We argue that our results point to a large  extension of the dominant and normative  paradigms  of  Asymptotic Freedom$/$IR colour confinement and   potential driven spontaneous symmetry breaking that have  long  ruled gauge theories. }  
 

 \newpage
\section{ Introduction} \hspace{01.4cm}

It has been  a longstanding dream\cite{wittenORaff}  to provide a mechanism for  dynamical generation of the Grand Unification scale from the low energy (i.e. Electro-weak)  data.  Asymptotic freedom(AF)  of the Grand Unified gauge coupling(s) has been a generally  unquestioned requirement for    acceptable  unification models.  Some years ago, motivated by the glaring Asymptotic strength(AS) of the couplings of the  phenomenologically satisfactory Minimal Supersymmetric Spin(10) GUT  model (MSGUT)\cite{aulmoh,msgut}, we proposed\cite{trmin,tas}  that this `defect'  is actually a signal from the model that it generates its own UV cutoff in the form of a Landau polar scale $\Lambda_{UV}$. This sets the scale  of    the (gaugino and other chiral)  condensates that form due to   strong coupling near  the  gauge Landau pole.  We show that  $\Lambda_{UV}$  is also  associated with  dynamical symmetry breaking   of the AS  gauge symmetry in the degenerate spontaneously broken gauge symmetry phases in the energy range below the Landau pole.     Inspired by their  defining role  in our proposal for  robust parameter counting ultra-minimal  AS Grand Unification we called  such condensates {\emph{pleromal}}. We were particularly enthused by the observation that -in contrast to, say, SuSy QCD-the nearly  exact Supersymmetry at the GUT scale implies that  the UV dynamics of the MSGUT, and other AS SuSy GUTs, are physically  the best justified and most realistic context  for the use of the  powerful methods \cite{seiberg,seibergduality} for analysing strongly coupled SuSy theories.

The phase structure  of  supersymmetric AS unifying models  which possess gauge Landau poles in the UV, often within an order of magnitude above $M_X$, must obviously be very  different from the well known AF scenarios \cite{georgidimo}. We do not agree with the common attitude that assumes AS theories  must be inconsistent and buries its head in the sand of a blind taboo against UV strong gauge coupling. It is a broadly applicable scientific truism that  in Nature there are no true infinities, but only naive idealisations.  UV gauge Landau poles should thus signal that a phase transition takes place and a new phase described by new field variables and gauge couplings comes into play. The canonical  example of QCD and its SuSy variants  has  served us  as a template.  On that analogy  we expect that near the gauge Landau pole  the  G-coloured particles (i.e. described by gauge variant fields) leave the  physical spectrum because their masses run to infinity and wavefunction renormalizations  run to singular values.  A new set of degrees of freedom analogous to the colour singlet Mesons,  Baryons, glueballs  and various {\emph{physical}} condensates of  QCD will be expected to describe the behaviour of the condensed   phase which should form in the strong coupling region.   In fact such candidate effective fields have long been identified in SuSy YM theories \cite{VY,seiberg} as being  holomorphic  gauge invariants formed from  the Chiral fields present in theory : which are known to parametrize the D-flat moduli space \cite{bucellaetal,affdineseiberg,lutytaylor} and are thus appropriate to describe the supersymmetric vacua. In analogy with the behaviour at strong coupling in the AF case one expects that a  supersymmetric Sigma model involving these gauge singlet but 't Hooft  anomaly matched `chiral moduli' fields  describes the UV phase\cite{tas}.

Be that as it may,  the  symmetry  restored, UV phase expected on general grounds such as RG flows   is not the subject of the investigations in this paper  beyond the fact that, again in direct analogy with SuSy QCD, we expect gauge invariant, physical, chiral  condensates to form in this phase. Once formed, being physical, we expect them to be a physical background  that all other phases of the theory must be consistent with. In particular  models based on full  renormalizable SYMH theories(i.e. with all massive modes retained)  are associated with the various degenerate broken symmetry  phases    at  scales below the UV condensation scale. These models  must all obey the infinite network of constraints between  SuSy vacuum condensates first  identified in the work of  Konishi and Shizuya \cite{konishi}  and  then greatly expanded in the work of Cachazo, Douglas, Seiberg and Witten(CDSW) \cite{GKA} and thereafter. 
   
In \cite{tas}  a toy AS GUT model with gauge group $SU(2)$ and a single symmetric chiral 5-plet was used to explore the  derivation of   symmetry breaking VEVs from the gauge singlet physical gaugino condensate.  The   use of the  Konishi Anomaly(KA)  \cite{konishi} allowed us to argue  that a VEV driven by the overall gauge singlet gaugino condensate might well develop. Shortly thereafter, a sophisticated and  powerful method  based on the  \emph{Generalized}  Konishi anomaly (GKA)  was invented\cite{GKA}  which allows   fully quantum and non-perturbative calculation of the condensates of  the ``Chiral Ring"  generators   $tr((W_\alpha W^\alpha)^n \Phi^m) | n=0,1, m \in Z_{\geq 0}$ in terms of the first few condensates.   Here $W_\alpha$  is the gaugino-field strength multiplet, $\Phi$ the   $N\times {\overline N}$  adjoint  multiplet of the Unitary gauge group and the trace is in the  N-dimensional fundamental since the adjoint has been written as an $N\times N$ matrix. Thereafter this method enjoyed a great vogue and was also extended \cite{seiberg2,GKA2}  by  the addition of   pairs of  fundamental (``quark" )  chiral multiplets  leading to either Higgs or ``pseudo-confining" vacua, or \cite{naculich}   by  other sets of chiral supermultiplets, such as (anti)symmetric representations : which provide more general rank breaking scenarios with several novel and non-trivial features in their  dynamics. 

In this paper   we   argue  that GKA techniques   allow  fully quantum and non-perturbative calculation of chiral VEVs in terms of assumed underlying gaugino condensates  in our \cite{trmin,tas}  asymptotically strong(AS) dynamical symmetry  breaking scenario. The scenario applies to a    vast   class  of  SuSy gauge theories   with   chiral multiplet $\Phi$  transforming as  $r\otimes \bar r$ : for {\emph{ any}}  representation $r$  of any gauge group G.  $\Phi$  may be      supplemented  by  representation pairs $\{Q,\bar Q\}; \{\Sig,\Sigb\} $ that can form   singlets  $\bar Q \cdot  \Phi^n \cdot Q,  \Sigb\cdot  \Phi^n \cdot \Sig$. Crucially,  $\Phi,\Sig,\Sigb$  can be  written as  matrices   (with rows and columns  labelled by indices running over the dimension $d(r)$  of  a general representation $r$ larger  than the fundamental : which we call the {\emph{base}} representation of the  model).  This allows the use\cite{GKA}   of resolvent methods to treat whole collections of condensates in terms of a single resolvent.   The basic idea of using a tensor product to define a matrix type representation can be extended to   gauge groups other than Unitary groups.  In particular it applies  to the product of spinorial representations of $Spin(N)$.   We note that by imposing trace constraints on the matrices $\Phi,\Sig,\Sigb$ that set some, or all but one, irreps contained in the direct products $r\times \bar r, r\times r$ etc.   to zero  one can try to  build models based on a subset of the irreps in $r\times {\bar{r}}$. However  this comes at a steep price of much more complicated loop equations and a larger set of condensate coding resolvents to be solved for. These problems have not been addressed yet. Moreover,
the normal motivation for choosing the smallest possible set of irreps of the gauge group is to reduce the number of free parameters in the Lagrangian  as well as to retain AF.  The latter reason is no longer applicable and using specific sub-irreps of $r\times \bar r$, $(r\times r)_{symm}$ etc. seems to lead to more, not less parameters. Hence in this paper we shall not attempt to extract sub-irreps.

The focus in the literature on GKA \cite{GKA,seiberg2,GKA2,CSW,naculich,brandhuber} has been on the restricted class of  Asymptotically Free(AF) i.e. IR strong     models and the  derivation of effective Wilsonian superpotentials to describe the (strongly coupled)  low energy theory.  In the GUT application case one rather wishes to know the  Higgs  vacuum expectation values(VEVs)   that must be substituted in the SYMH Lagrangian to derive the effective {\emph{perturbative}}  spontaneously broken GUT(i.e. with the dynamical symmetry breaking sue to strong coupling at the UV Landau pole  incorporated into the parameters of the effective SYMH)  and therefrom its effective supersymmetric light mode theory(a.k.a ``MSSM''). Thus one   needs a definition for the {\it{quantum  VEVs}}  and associated ``equivalent  quantum superpotential'' $W^{(q)}(\Phi, \lambda,v_i^{(q)}) $.  This is a (novel) supplementary  superpotential, which we introduce for  the   specific purpose of coding in the ``quantum VEVs''  derived via GKA based  analysis,   into a generalisation of  the tree superpotential $W (\Phi, \lambda...) $  so as  to  define  a consistent SYMH Lagrangian based starting point for the spectrum analysis of the GUT in its  spontaneously broken phases.  The  extrema   of  $W^{(q)}$ are   the quantum corrected  VEVs $v_i^{(q)}$ computed by the non-perturbative GKA formalism involving solutions for, and contour integrals of,  quantum resolvents i.e. by non potential extremisation methods.   

Our focus here is  to explore {\emph{the generation of the GUT scale  chiral VEVs, in each of the possible   symmetry broken phases,    by  non-perturbative and fully quantum dimensional transmutation}}, especially  when the mass parameters in the superpotential are absent or negligible.   We emphasise that the parameters describing (a few) irreducible {\emph{gauge singlet}} physical condensates are to be   deduced {\emph{ in principle}}  from the non-perturbative  dynamics of the UV symmetry restored phase using, say, Lattice methods.  In  practice, however, they  are simply  unknown dimensionful input parameters playing a VEV determining role similar to the mass parameters of standard SuSy GUT superpotentials.   
If one accepts that Asymptotic Strength should result  in  an underlying  physical G-singlet gaugino condensation then  in the spontaneously broken phases relevant at lower energies the gaugino condensate `fractionates' into different parts corresponding to a division of the G -gauginos into those associated with  the little group $H$ and cosets $G/H$, while maintaining the overall condensate value.  
 
 In this work  we shall outline the generic features of our proposal applicable to a large class of models with arbitrary gauge group, noting in conclusion only the principal features of its  application to a  realistic SO(10) model. In short, we suggest a   quite  novel approach to the hoary problem of GUT symmetry breaking which realises the old dream of  symmetry breaking scale determination by dimensional transmutation, not in a engineered  perturbative model\cite{wittenORaff,CW}, but generically and robustly in an infinite class of  models of a type hitherto largely neglected.   Our scenario   presents a plausible picture and provides useful and novel calculation  techniques for analysing hybrid theories where the deep UV  phase G-singlet gaugino condensates   drive  dynamical symmetry breaking in the lower energy phases. It   should thus  help to free AS theories from the limbo to which they have hitherto been banished by a taboo that assumes they are automatically logically inconsistent simply because no method had been found to calculate anything interesting about them. If so it   may  have much wider implications and ramifications for gauge theories as a whole. 

In Section 2.1  we first review the GKA analysis of CDSW  applied to  a very restricted type of AF SuSy YMH models with $U(N)$ gauge group and adjoint Higgs irrep.  Then in Section 2.2 we briefly  discuss    generic features of the effective theories associated with gaugino condensates in AF vs AS SYMH models   to provide a context for the calculations  we perform. In Section 2.3,  we  discuss how the techniques of \cite{GKA} for Adjoint Multiplets extend to an infinite class of ``Generalized Adjoint multiplet type"(GAM) models based on a  notion of base-$r$ tensors  transforming as direct products  $r\times \bar r$ of the gauge group G. In Section 2.4 we define the quantum VEVs and superpotential $v^{(q)}_i,W^{(q)}$.   In Section 3.1 we give a simple example of this extension with gauge group SU(3) and base representation $r=(3\times 3)_{symm}$ and illustrate calculations with some numerical results. In Section 3.2 we   illustrate  `gaugino condensate fractionation' using the example. In Section 4. we discuss how rank breaking may be implemented by introducing additional $(r,\bar r)$ or $(r\times r,\bar r\times \bar r)$ pairs of Chiral supermultiplets whose VEVs reduce the rank of the little group $H$. In Section 5. we outline a realistic $Spin(10)$ GUT model whose  base representation is the $16$  dimensional chiral spinor of $Spin(10)$. In Section 6. we discuss our results and the outlook for further work  from a general view point.

 \section{Generalized Adjoint models}
\subsection{Generalized Konishi Anomaly and CDSW  formalism for matrix form  representations }    We begin with a short summary of the basic work \cite{GKA}  on which we rely for our calculations. This work confines itself to consideration of the standard Asymptotically free  SYMH models  which condense in the Infrared.  In  a  supersymmetry preserving  vacuum  of a super-Yang Mills theory with gauge group G coupled to a Chiral multiplet $\Phi$ in an arbitrary representation $R$ of the gauge group, and   with superpotential ${{W}}(\Phi)$,   the GKA implies\cite{GKA,naculich,brandhuber}  the following relation for  condensates of chiral gauge invariants formed from the Gaugino-Field strength  Weyl spinor  chiral multiplet  $W_\alpha^A, A=1...dim(G), \alpha=1,2 $ and the chiral fields  $\Phi_I, I=1.... dim(R)$ :
 \bea \langle f_I  {\frac{ \partial { {W}} }{\partial \Phi_I}} \rangle = -{\frac{1}{32 \pi^2}}   \langle W_\alpha^A  W^{\alpha B}  {{\cal{M}}^A}_I^J { {\cal{M}}^B}_J^K  {\frac{ \partial { {f( W_\alpha,\Phi)_K}} }{\partial \Phi_I}}  \rangle  \label{GKA}  \quad ; \quad 
 <W^{\alpha  A}  {{\cal{M}}^A}_I^J  \Phi_J>  &=& 0 \eea
 Here $f( W_\alpha,\Phi)_I $ is an arbitrary chiral variation of the field $\Phi$ in the representation $R$ with generators ${{\cal{M}}^A}$. Repeated indices $I,J,K $ are summed over $dim(R)$  values.  The important constraint  equation whereby the matrix $W_\alpha$  acting on the ``vector form'' of  the general representation $\Phi$ is equivalent to zero in the ``Chiral Ring" \cite{GKA}  is frequently used in simplifying expressions  for chiral expectation values. Henceforth, we drop angular brackets to indicate expectation values of operators since that is all we ever  consider.
  
  For the case where $\Phi$ transforms as the  traceful  adjoint of $U(N)$ the method of \cite{GKA} allowed a complete solution for the generators of the Chiral ring  of gauge invariants $t_{n,m} \sim  tr((W_\alpha W^\alpha)^n \Phi^m) ( n=0,1;  m \in Z_{\geq 0}$ ).  The solution proceeds  by solving for the generating functions of the Chiral Ring generators defined as  the resolvents   $tr (W_\alpha W^\alpha)^n (z-\Phi)^{-1}$ . These have   obvious expansions  as power series, valid for large $z$, whose coefficients are the Chiral Ring Generators $t_{n,m}$.  In \cite{seiberg2,GKA2}  the  extension to the case with additional (``quark") superfield pairs $\bar Q, Q$ which can form a singlet with $\Phi$  as $\bar Q\cdot \Phi^n \cdot Q $  and  in \cite{naculich}   the similar case of the Adjoint with conjugate  pairs of   (anti)symmetric  representations $(\Sigb,\Sig)$  are resolved. These additional models allow consideration of rank breaking supersymmetric Higgs vacua not available with just an adjoint. Thus
   \bea  R(z) &=&\kappa \,  tr ( W_\alpha   W^\alpha (z-\Phi)^{-1})  \equiv\sum_{n=0}^{\infty}  {\frac{R_n}{z^{n+1}} }\equiv 
    \kappa  \sum_{n=0}^{\infty}  {\frac{tr(\Phi^n   W_\alpha W^\alpha)}{z^{n+1}}}    \nnu
   T(z) &=& tr  (z-\Phi)^{-1} \equiv \sum_{n=0}^{\infty}  {\frac{T_n}{z^{n+1}} }\qquad ; \qquad \kappa \equiv -{\frac{1}{32 \pi^2}}  \eea  
   $tr$ is taken  so that the matrix indices run over  $1...d(r)=N$.
  The semi classical vacuum of the  model is defined by distributing the critical values  $a_i  (i=1..  n  | { {W}}'(a_i) =0)$    of the superpotential function ${ {W}}(z) $ (degree $n+1$)  over the diagonal slots  of $\Phi$ and setting the off-diagonal elements to zero (this ensures minimisation of the D-term contributions to the potential  via  $D^A(\Phi,\Phi^*)\sim [\Phi,\Phi^\dagger]=0$).  One can extract various interesting quantities, such as the number $N_i$  of times a critical point  $a_i$ is repeated or the value  of   gauge invariant  chiral ring generators,  via  integrals  of $ z^m \{R(z),T(z)\} \, $ etc.  around suitable contours $C_i$.  
 
  \subsection{Generics of SYMH condensates }
 Super-QCD without quarks   is  AF  and so has an IR  Landau pole and associated RG-invariant scale $\Lambda_{SQCD}$. In the seminal work of Veneziano and Yankielowicz(VY)  \cite{VY}  gaugino  condensates consistent with the expected unbroken supersymmetry  were found. In contrast to QCD, no condensates (like {\hbox{$ \alpha_3<G_{\mu\nu}^A  G^{A\mu\nu}>\sim \Lambda_{QCD}^4 $}})    leading to a contribution to the vacuum energy  are permitted by unbroken supersymmetry.  On the other hand a  condensate for  the gaugino bilinear $<\lambda_\alpha^A  \lambda^{\alpha A} > \simeq \Lambda^3$ is chiral (i.e. F-type)  and does not break SuSy.   SQCD with $0<N_f< N_c$ quarks has no stable vacuum but for larger values of $N_f$ exhibits \cite{seibergduality} interesting phenomena like a   `conformal window' for $3 N_c > N_f > 3 N_c/2$ where the theory has  a  superconformal fixed point  and is in  an `interacting non-Abelian Coulomb phase'. Our basic assumption when considering the inverse i.e. AS case is that  gaugino condensates continue to be formed due to the strong coupling dynamics associated with the (UV) Landau pole.

 In  IR-strong SuSy theories   the unbroken symmetry is set already at high energies where the gauge couplings of the little group  $H=\Pi_i U(N_i) $  left unbroken after $U(N)\rightarrow H$  breaking are still perturbative.   The $N_i$ refer to the VEV repetition numbers   on the $\Phi$ VEV diagonal. The   GKA analysis showed that the vacuum structure of the fully quantum theory is controlled by branch cuts arising from bifurcations of  the semi-classical critical points  $a_i$   due to  quantum corrections.  The branch-cut network has a physical meaning in its own right   since the  `quantum superpotential derivative'  which defines the branch cuts  makes no reference to the semi-classical   critical points. The  gaugino condensate and a few other  low lying condensates involving the adjoint supermultiplet $\Phi$   determine the `quantum superpotential derivative'. Consistency implies  the branch points   coalesce into the semi-classical critical  points when the quantum corrections coded in the driving condensates  are set to zero by hand.  In the IR the non-Abelian gauge  couplings  become strong and eventually gaugino condensates $<S_i> \sim W_{\alpha}^{A_i} W^{\alpha A_i}, A_i= 1...N_i^2 $  form in each of  subgroups $U(N_i)$.  We take over this   picture {\emph{mutatis mutandis}}.    The IR   effective theory is now  the spontaneously broken SYMH theory  and not the confined gauge theory of $\Pi_i U(N_i)$.  Our objective is to calculate and encode the symmetry breaking VEVs into a suitable supersymmetric SYMH Lagrangian  rather than  to calculate the VY effective potentials.  The special techniques available in supersymmetric gauge theories allow  the extraction of significant information regarding  the whole system of condensates. Our aim is firstly to provide a robust underpinning of automatic  UV condensation for   GUT SSB. Secondly we aim  to  reduce the overall number of free parameters of the massive perturbative GUT model which gives rise to the effective low energy theory. Higgs VEVs develop and are determined by the UV Landau polar scale  {\emph{even without any input mass parameters in the superpotential}}.   Thus it is justified to claim that, in sharp contrast to \cite{CW},  Dimensional Transmutation   has  emerged robustly from the UV strong coupling dynamics.  
 
  The contour integrals    used to extract the fully  quantum GUT symmetry breaking  VEVs  in terms of the input condensate parameters(see below for details)   are a far cry from the  habitual semi-classical minimisation of a scalar field potential.   Once we have them in hand we encode them in the form of the parameters of a spontaneously broken Supersymmetric gauge theory of the standard type specified by a ``equivalent  quantum superpotential " ($W^{(q)}$)  with  a new set of massive parameters containing information about the effect of the non-perturbative condensates for the perturbative GUT theory.   
     
  \subsection{GKA techniques for  Generalized Adjoint Multiplets (GAMs)}  
    
  Our basic observation is that the  equations for resolvents derived via the GKA in \cite{GKA} also hold for the Generalized Adjoint Multiplet(GAM)   type field transforming as $r\times \bar r$ for {\emph{any }}  representation $r$ of any simple/semisimple gauge group  G with the {\emph{sole}} replacement of the trace ($tr$) in the fundamental of $SU(N)$ by the trace ($Tr$)   in the representation $r$  of G. Note that $d(r)=N$ for the model of \cite{GKA}  but their results generalise easily using  the expanded notion of an ``base-$r$ Adjoint", written as a $d(r) \times d(r)$ matrix, that transforms as $r\times \bar r$ rather than as $N\times \bar N$. In other words, under a gauge transformation
  \be \Phi'= U_{d(r)} \cdot \Phi \cdot {U_{d(r)} }^\dagger \ee
  where $U_{d(r)} $ are $d(r)\times d(r)$ dimensional Unitary matrices in the representation $r$ of $G$.  The generators can thus be written as \bea {\cal M}^A_{r\times{\bar r}} &=& T^A_{(r)} \times {\cal{I}}_{d(r)} + {\cal{I}}_{d(r)}   \times   T ^A_{(\bar r)}= T^A_{(r)} \times {\cal{I}}_{d(r)} - {\cal{I}}_{d(r)}   \times   (T^A_{(r)})^T \eea   and $T^A_{(r)}$ are Hermitian Generators in the representation $r$. Then it is easy to rewrite eqn(1) in the form it was derived in \cite{GKA}  with the generalization $T^A_{fund}\rightarrow T^A_{(r)}$ : 
  
  \bea < f_{ij} {\frac{\partial W(\Phi)}{\partial \Phi_{ik}}}> &=& \kappa <W^A_\alpha W^{B\alpha} {\frac{\partial}{\partial \Phi_{ij}}} [T^A_{(r)},[T^B_{(r)} ,f(W_\alpha,\Phi)]]_{ij} > \eea
   In general,  $r\times \bar r$ contains one or more  irreps of G, {\emph{besides}} the singlet and the adjoint always present, and such representations   carry large $S_2(R) >> 3 C_2(G)$. By constraining the matrix $\Phi$ so as to single out one or more irreps (which are still  AS)  one can work in terms of irreps of G. For example one can remove  the singlet and adjoint in $r \times {\bar r}$ by imposing $Tr(\Phi)= Tr(T^A \Phi)=0$ and   likewise for $\delta \Phi$.  The required resolvent formalism will be quite complex and is not yet available.   
      
 We define  
\bea  I[C,F(z)] &\equiv&  {\frac{1}{2 \pi i}} \oint_C  dz F(z)\nnu 
N_i &=&  I[C_i,T(z)]  \nnu
 R_0 &=& I[C_\infty,R(z)] =\kappa S_2(r) W_\alpha^A W^{\alpha A} =2 S_2(r)  {\cal{S}}    \nnu
 R_n &=& I[C_\infty,z^n R(z)]\qquad ; \qquad 
 T_k = I[C_\infty, z^k T(z)]    \eea
  Where ${\cal{S}}$ is the gaugino condensate and $S_2(r)$ the index of the representation $r$ :\hfil\break  $Tr({ {T^A_{(r)}}} {{T^B_{(r)}}})=\delta^{AB} S_2(r)$.  It is obvious that 
$ v_i^{cl} =a_i  =  I[{C_i}, z \,  T(z) ] /N_i$  extracts the VEV $v_i= \Phi_{ii}$   of an $N_i$-fold replicated  diagonal $\Phi$ component from the  semiclassical    $T(z)$ when $C_i$ encloses the critical point $a_i$. {\emph{This  motivates the VEV  definition in the quantum corrected case when $C_i$   encircles  the branch cut connecting  pairs of branch points which are bifurcates of the semi-classical critical points split under the influence of quantum corrections\cite{GKA} }} (see  eqn(\ref{VEVdef}) below).     

  In the presence of quantum corrections the   GKA method  of \cite{GKA} solves for the generating functions  $R(z)$   using the position independence and complete  factorizability\cite{GKA}  of  Chiral ring operator correlators  to reduce the  GKA (for the case where  $\delta \Phi =\kappa  W_\alpha W^{\alpha}  (z-\Phi)^{-1}$)    to   a quadratic equation    for $R(z)$  :    
\be R(z)^2 -{ {W}}'(z) R(z) -{\frac{1}{4}} f(z)=0 \label{ResR}\ee
 where $f(z) $ is a  degree $n-1$ polynomial which is  determined by the driving condensates $R_{0,1,...n-1}$  and   ${ {W}}(z)$ is just the superpotential (of degree $n+1$) as a function of $z$.   Since our entire focus is on the relevance to renormalizable GUTs, we shall only  consider cubic superpotentials ($W(z) =\lambda z^3/3 +m\, z^2/2 +\mu^2 z$).  Then  $f(z)$ is a linear function  $f(z)= f_0 +f_1 z  = -4 \lambda (R_1+z R_0)-4 m R_0$.   The coefficients  $R_0,R_1$  are not determined by the GKA  and should be regarded as dynamical moduli of the vacuum manifold  of the theory  which are to be determined by an appropriate numerical investigation of   gaugino condensation at strong coupling. Some information about the main contribution to $R_1/R_0^{4/3}$ can however be gleaned by surveying  the GKA constraints numerically. 
  
  In the AF case of   SuSy YM with Adjoint Chiral Higgs the gauge coupling runs to a Landau pole {\emph{in the infrared}}   at a (RG invariant) scale  $\Lambda$ approximated by the one loop value (exact for the Wilsonian gauge  coupling)  \be \Lambda   =  \mu  e^{\frac{8 \pi}{b_0 g^2(\mu)}} \label{Lambda}\ee 
  where $b_0=-2 N$ for the adjoint-SYM.
  There are good arguments\cite{VY}  to support the conjecture that  strong coupling causes the development of a gauge singlet, RG invariant and physical gaugino condensate at this scale \be {\cal{S}}\equiv  {\frac{\kappa}{2}} W_\alpha^A W^{\alpha A} = b \Lambda^3 \ee Where the constant  $b$  is scheme dependent and may be chosen to be 1\cite{GKA}.

  Our core assumption is that physical gaugino condensation also occurs when the gauge coupling runs to a Landau pole {\emph{in the ultraviolet}},  at a scale $\Lambda_{UV}$ still given by (\ref{Lambda}),  but for the case $b_0(R)=S_2(R)-3 C_2(G) >0$.   Note that $S_2(r\times \bar r)=2 \, d(r)\, S_2(r)$, which grows fast with $d(r)$,  so that $b_0(r\times\bar r) >>0$  for most base representations $r$.  We   deduce, via the GKA relations and consistency conditions,  the symmetry breaking quantum  VEVs of $\Phi$  that define the effective low energy theory as functions of the basic gaugino condensates and the superpotential parameters. The analysis is performed at a scale where all the degrees of freedom of the Super YMH theory are retained with  the RG invariant gaugino condensates as a given background.  Scale dependent quantities such as superpotential parameters and VEVs should be regarded as    defined at such an intermediate scale where the gauge and superpotential couplings are still perturbative. 
    
   By assumption, a  G-singlet physical gaugino condensate ${\cal{S}}$ for the group as a whole  develops  in the {\emph{ultraviolet}}. This gaugino condensate then requires (via the GKA and its solution)  that the fields develop VEVs (calculated below) which break the gauge symmetry in a manner dictated by the placement of these quantum corrected VEVs  $v_i^{(q)}$ on the diagonal of the $\Phi$ VEV {\emph{in any way(i.e. with any set of $\{N_i\}$)   we choose}} : thus defining a variety of degenerate spontaneously broken supersymmetric  phases. This placement of VEVs  determines the little group  $H$ in practice. The different gaugino bilinears condense in a pattern determined by the VEV placement.The assumption that one may choose the  $N_i$  (subject to $d(r)=\sum_i  N_i$ )  to be fixed integers  is an important constraint on dynamical symmetry breaking.    
  
       The solution of the GKA equation for $R(z)$ is 
\bea R(z) &=& {\frac{1}{2}} (  { {W}}' (z) -\sqrt{ {{ {W}}'(z) }^2 +f(z)} )\equiv {\frac{1}{2}} (  { {W}}' (z) -y(z) ) \label{Rqdef}\nnu
y^2(z)&=& W'(z)^2 +f(z) \quad ; \quad    f(z)  \equiv  4\,  \kappa \, Tr ( W_\alpha W^\alpha({ {W}}'(\Phi) -{ {W}}'(z))   (z-\Phi)^{-1})      \eea
The GKA equation obtained for $\delta\Phi= (z-\Phi)^{-1}$ yields 
\bea (2 R(z) -W'(z)) T(z) -{\frac{c(z)}{4}}&=&0 \label{Tloop} \nnu
 4 \, Tr ( ({ {W}}'(\Phi) -{ {W}}'(z))   (z-\Phi)^{-1})  &\equiv & c(z) \eea   Like $f(z)$,     $c(z)$  is  a polynomial  of  degree $n-1$.   Thus 
  \bea T(z) = -{\frac{1}{4}} {\frac{c(z)} { \sqrt{ {W'(z)}^2 +f(z)}}}\equiv -{\frac{ c(z)}{4\, y(z)}} \label{Tqdef}\eea
 For a cubic superpotential ($n=2$)   the  expansion of $T(z)$ for large $z$   and the above solution for $T(z)$, yields  $c_0=-4 \lambda T_1 -4 m \, d(r), c_1=-4 \lambda \, d(r), T_0 \equiv d(r)$. 
The square root  of the   polynomial $y^2(z) =W'(z)^2 +f(z)$   encodes the `quantum  superpotential  derivative', which is distinct from  the classical one if $R_{0,1} \neq 0 $. The higher  coefficients $R_n, T_n$ which are not present in the solutions  $y(z), T(z)$ are determined in terms of $R_{0,1....n-1}, T_{0,1...n-1}$  by the GKA relations. For the cubic  case ($n=2$)  the  unknowns are thus $R_{0,1},T_1$. The gaugino  condensate   $R_0=2 S_2(R) {\cal{S}} = 2 S_2(R) \Lambda^3$ sets the scale of all other condensates and is estimated directly from the running of the perturbative gauge coupling in the full theory. Thus, for numerical work,  we can conveniently rescale  all our expectation values and integration variables to dimensionless forms using units of $R_0^{1/3}$.

  For a  cubic superpotential  $R(z),T(z)$ the first few dependent $R_n,T_n$ are :
  \bea
 R_2 &=& - {\frac{\mu^2 R_0 - m R_1}{\lambda \, }} \quad ; \quad
 R_3 =  {\frac{1}{\lambda \, ^2}} (m \mu^2 R_0 + m^2 R_1 + R_0^2 \lambda \,  - 
   \mu^2 R_1 \lambda \, ) \nnu   T_2 &=& -{\frac{1}{\lambda \, ^2}} ( d(r) \mu^2 -  m \, T_1)  \quad ; \quad
 T_3  =  {\frac{1}{\lambda^2}} (d(r)\, m\, \mu^2 + m^2 T_1 + 2  \lambda \,  d(r) R_0 - 
   \lambda \,  \mu^2 T_1 ) \nnu
     R_4 &=&{\frac{1}{\lambda^3}} (-m^2 \mu^2 R_0 - m^3 R_1 + \lambda \,  \mu^4 R_0   - 
    \lambda \,  m R_0^2   + 2  \lambda \,  m \mu^2 R_1   + \lambda^2\,   2 R_0 R_1 ) \\ 
T_4 &=&  {\frac{1}{\lambda^3}}(-d(r) m^2 \mu^2 - m^3 T_1 +  \lambda \,  d(r) \mu^4   - 
   2  \lambda \,  d(r) m R_0  + 2  \lambda \,  m \mu^2 T_1   + 
   2  \lambda^2 \,d(r) R_1  + 2  \lambda^2\, R_0 T_1 )\nonumber  \eea

The classically superconformal case  where $m=\mu=0$, i.e. the tree level superpotential is free of any mass parameters,   is particularly simple and interesting :
 \bea R_2 &=& 0\, ;\,  R_3 = {\frac{R_0^2}{\lambda \, } }\,\, ;\,\,  T_2 = 0 \,\,  ;\,\,
 T_3  =  {\frac{2 d(r)  R_0}{\lambda \, }}\,\, ;\,\, 
 R_4 = {\frac{2 R_0 R_1}{\lambda \, }} \nnu 
 R_5 &=& {\frac{R_1^2}{\lambda \, }}\qquad ;\quad 
 T_4 = {\frac{2 (d(r) R_1 + R_0 T_1)}{ \lambda \, }}\quad ;\quad  T_5 = {\frac{2 R_1 T_1}{\lambda \, }} \eea

If, restoring angular brackets for a moment,  we separate $\Phi(x)=<\Phi> +\tilde\Phi(x)  $ where $\tilde{\Phi}$ has zero VEV, we see that $T_1= Tr{{<\Phi>}}= \sum N_i v^{(q)}_i$ but \bea T_2&=&<Tr\,\Phi(x)^2>=<Tr\,\Phi(0)^2> =Tr<\Phi>^2 + <Tr({\tilde\Phi(0)}^2)>=0\,\nnu \Rightarrow \, \sum N_i {v^{(q)2}_i}   &=&-<Tr({\tilde\Phi(0)}^2) >\eea   This sort of interplay between non-gauge invariant quantum correlators  and non-gauge invariant VEVs  can  yield  gauge invariant  resolvent coefficients $R_n,T_n$. Thus although we shall use   vacuum expectation values deducible from $T(z)$ to calculate masses and define a  effective theory with spontaneously broken gauge group at low energies, we must keep in mind that due to the strongly non-perturbative physics at high energies the GKA relations imply a host of strong constraints on higher order chiral correlators whose implications for the effective theory remain to be explained. The phenomenological implications of these  constraints for correlators involving quantum fields which are known to describe light particles, like SM fields, are not clear to us. We  regard this hybrid of a perturbative effective theory  with strong correlations due to underlying microscopic  strong coupling as one of the most puzzling and  intriguing implications of our results : which may provide fresh  insight on how to think about vacua arising non-perturbatively  in strongly correlated systems.

 We argue below that consistency of the GKA relations with the choice of $N_i$ also determines $T_1$ in terms of contour integrals involving $y(z)$   and the  integer repetition numbers   $N_i $ of the VEVs $v_i^{(q)}, i=1..n$ on the diagonal  of $\Phi$ which, following \cite{GKA},  we assume to be free inputs stable against quantum deformation. This only leaves $R_0$ and $R_1 = \kappa \,Tr \, \Phi  W_\alpha W^\alpha  $ undetermined.  We shall see below that the GKA equations offer insight and an estimate  even for $R_1$  : at least in cases where the effects of the condensation  are primarily encoded in the quantum VEVs and gaugino condensates.
  
\subsection{Determination of Quantum VEVs $v^{(q)}_i$}
The essence of the analysis of \cite{GKA}   in the AF  case is that the influence of   gaugino condensation modifies  the polynomial $y^2(z)$ such that its zeros are bifurcates  $a^{\pm (q)}_i $  of the critical points $a_i\,  (i=1...n | W'(a_i)=0) $. These pairs    are    branch points of $y(z)$ and are connected by branch cuts.    They merge when the quantum condensates $R_0,R_1$ are sent to $0$ and $y(z)$ reverts to its classical value $W'(z))$.  The polynomial $y^2(x)$  defines a two sheeted Riemann surface(of genus 1  when $n=2$ since $y^2$ is then quartic). The contours $C_i$ enclosing the semi-classical  critical points  $a_i$ now  become contours $A_i$ enclosing  branch cuts running between the corresponding pairs of branch points $a^{-(q)}_i,a ^ {+(q)}_i$. Thus  $\{A_i,i=1..n\}$ become\cite{GKA}   a (once redundant) basis for the A-cycles  of the Riemann surface defined by $y^2(z)$.  The definitions of the sub-condensates in the quantum case  involve integrals around the A-cycles :
  \bea       R_{0i}  &\equiv& {\frac{1}{2 \pi i}} \oint_{A_i} dz \,  R(z) =-{\frac{1}{4 \pi i}} \oint_{A_i} dz \,  y(z)=-{\frac{1}{2}}   I[A_i,y(z)]\nnu
     R_{1i}  &\equiv& {\frac{1}{2 \pi i}} \oint_{A_i} dz \, z \,  R(z)=  -{\frac{1}{4 \pi i}} \oint_{A_i} dz \, z \,  y(z)= -{\frac{1}{2}}   I[A_i, z\, y(z)]  \eea
  Most crucially, we propose that the quantum VEVs should be defined as 
     \bea 
      v_i ^{(q)} &\equiv &  {\frac{1}{2  \pi i N_i}} \oint_{A_i} dz  \,  z \, T(z) =             -{\frac{1}{8  \pi i N_i}} \oint_{A_i} dz  \, {\frac {z \, c(z)} {y(z)} } 
       \label{VEVdef} \eea  
       It is clear that as $ f(z) \rightarrow 0$  the VEVs approach their classical values $v^{(q)}_i \rightarrow a_i$.   To our knowledge, though natural and obvious, it has not been explicitly considered earlier since in \cite{GKA} and thereafter the emphasis is on gauge invariant specification of the SSB in the UV via $<Tr \Phi^n >$ rather than deriving VEVs whose substitution  will produce   Lagrangians  for the spontaneously broken  phases of the GUT.  As emphasised in  \cite{GKA}, the sets of  integers  $N_i$  which  invariantly specify the little group $H$  should not change even when evaluated (using the solution for the resolvents) as 
\bea N_i &\equiv&  {\frac{1}{2  \pi i  }} \oint_{A_i} dz  \,   T(z)        = -{\frac{1}{8 \pi i}} \oint_{A_i}\,  dz\, {\frac{c(z)}{y(z)}}  \label{Nidef} \eea

In contrast to the $r=N,\Phi\sim N\times {\overline{N}}$ case studied in the AF case in \cite{GKA}, the sub-condensates $R_{0i}$ correspond {\emph{not}}  to unbroken  subgroup factors'  gaugino condensates  but instead to certain combinations of the gauginos of the unbroken gauge  sub-group H  and the  $G/H$ coset gaugino condensates. The  combination relevant for $R_{0i}$  can be identified by  evaluating  $Tr T^A_{(r)}  T^B_{(r)}  {\cal{I}}_{d(r)}^i$ where  in ${\cal{I}}_{d(r)}^i$ only the unit diagonal elements in the sector corresponding to the cycle $A_i$ are retained and the others are set to zero. This is illustrated explicitly with an example in the  next section. 
 The consistency of these contour integrals   with the Laurent expansion even in the quantum corrected case is ensured by the fact that
 \bea \sum_i \oint_{A_i}   z^n (y(z))^{\pm m} \, dz  = \oint_{C_{\infty} }   z^n (y(z))^{\pm m} dz   = \oint_{C_{z'=0} } \, dz'  z'^{-(n +2)} (y(1/z'))^{\pm m}   \eea  
 since the region enclosed by the  curves  $\cup_j   A_j  \cup (-C_{\infty})$ is free of  singularities or branch cuts.
 
 For   $n=2$,  $T(z)=(d(r) (z  \lambda  +m)  +\lambda\,  T_1) y^{-1}$, and   we can solve the  definition of  $N_1 $ to obtain   a consistency condition of the assumption\cite{GKA} that the value of the integrals giving $N_i$ around the critical points $a_i$ do not change under quantum corrections.   From  eqn(\ref{Nidef})we get  
   \bea T_1 &= & {\frac{  2 \pi i N_1-\oint  dz \, (\lambda \, d(r) z +m \, d(r)) y^{-1} }   {\oint  dz\,\lambda \,  y^{-1}} } \eea
  The equation for $N_2$ gives nothing fresh because of the complementarity of the integrals around $C_{\infty} $ and the  union of the A cycles.   When $n>2$  we can similarly obtain  equations for $T_{1}....T_{n-1}$ by using the definitions of $N_1...N_{n-1}$ and solving the $n-1$  linear equations for the $T_i$.

In the pure $r\times \bar r$ case,  analysis of the GKA equations also yields insight and constraints upon the form of $R_1$.  Closed form evaluation of the elliptic or hyper-elliptic type integrals involved  is difficult  since $\lambda, R_1,a^{\pm,(q)}_i$ are, in general,  complex.   Numerical evaluation of  expressions for $T_1,v_i^{(q)},R_{0i},R_{1i}$ (in units of $ R_0^{1/3}  \sim \Lambda$)  for  a range of values of $\lambda,R_1$  and a choice of  $N_i$  is straightforward. An important class of possible (even necessary, if the dynamics is to support  a picture where the main effect of the condensation for defining the  effective perturbative theory is coded in quantum  vacuum expectation values $v_i^{(q)}$ of the  field $\Phi$ )  solutions are those where   $\sum_j   R_{0j} v_j^{(q)} $ closely approximates $ R_1=\sum_j  R_{1j}$ i.e. for which the dimensionless semi-classicality parameter is small  : \be  \delta_{SC}(R_1)\equiv |   {\frac{ R_{1}-\sum_j^n  R_{0j} v_j }{R_1}} | <<1   \label{deltaSC}\ee   While we cannot calculate the dimensionless condensate ${\hat R}_1\equiv R_1/R_0^{4/3}$  we do  find that  there are regions of  the dimensionless parameter space  $(\lambda, {\hat R}_1)\in {\mathbf{C}}^2$  where   $\delta_{SC}  $ is  very small. Heuristically speaking, such regions  in the parameter space are   candidate vacua where the gaugino condensate and the quantum corrected VEVs encode most of the  dynamical information relevant for defining the effective perturbative theory below the scale of dynamical gauge symmetry breaking.  We emphasise that  $\delta_{SC}$ is merely a numerical measure useful in understanding the types of solutions obtained while numerically scanning the parameter space of the theory (including the values of the non-perturbative physical condensates such as $R_{0,1}$)  while evaluating the   quantum VEVs.  It is {\emph{ not}}  any sort of constraint imposed on the supersymmetric dynamics to ensure its consistency, but merely a parameter that indicates that the quantum dynamics is in some sense close  to semiclassical   when it is small.

\subsubsection{The Equivalent Quantum Superpotential $  W^{(q)} (z)$}
 We  propose  that the  VEVs  $v_i^{(q)}$  be used to   define a consistent  effective  Lagrangian  for calculating mass spectra in  the spontaneously broken theory  by modifying  the semiclassical  superpotential  ${ {W}}(z)$  to a quantum modified or effective superpotential $  W^{(q)} (z)$ by changing  the  coefficients in ${ {W}}(z)$ so that the VEVs  obtained are zeros of ${W'^{(q)\,} } (z)$ i.e.  ${W'^{(q)\,} } (v^{(q)}_i)=0$.  Thus, for example for a cubic superpotential,   we take  :
\bea W^{(q)} (z) &\equiv & \mu_{(q)}^2   z +{\frac{m_{(q)} }{2}} z^2 +{\frac{\lambda}{3}} z^3\ \nnu 
 \mu_{(q)}^2 &\equiv&  \lambda v^{(q)}_+ v^{(q)}_-   \qquad ; \qquad m_{(q)} \equiv -\lambda (v^{(q)}_+  +v^{(q)}_-){\label{Wqdef}} \eea
Classically $v_\pm={\frac{-m\pm \sqrt{m^2 -4 \mu^2 \lambda }}{2 \lambda}}$ and one recovers the original superpotential.
This proposal   has the virtue that the cubic coupling describing the interactions has not been modified  and the changes made are only in the  super-renormalizable couplings.  Use of such a superpotential will ensure that the super-Higgs effect and spectrum  calculations using $v^{(q)}_i$ are  consistent.

For evaluating the contour integrals around the $A_i$ cycles   we should first define the  square root branched function $y(z)$ to lie unambiguously on the first sheet  :
 \be y(z) \equiv \lambda   \prod_{i=1}^{2n} |(z-z_i) |^{\frac{1}{2}} \prod_{i=1}^{ n}  e^{{\frac{i }{2}} (\theta ({\frac{z -z_{2 i-1} }{z_{2i}-z_{2 i-1}}} )       +\theta ({\frac{z -z_{2 i } }{z_{2i}-z_{2 i-1}}} ) )}  
 \ee
 where $\theta(z)\in(-\pi,\pi]$ is the quadrant wise correct polar angle of the complex number $z$.
 Then the integral over the  $A_i$-cycle  which encloses the branch cut running from $z_{2i}$ to $z_{2i-1}$ is 
  achieved by (P.V. denotes principal value and the function $g(z)$ should be such that the contribution from the end  circles around the branch points is zero)     \be \oint_{A_i} dz \, g(z) = 2 \, \, P.V. \int_0^1 \, dx\,  ( z_{2i-1}-z_{2 i}) g( x( z_{2i-1}-z_{2 i}) +z_{2 i})  \ee
 
   If the dynamical behaviour supports the emergence of an effective   spontaneously broken perturbative theory  then  we expect that  $\sum_j R_{0j} v_j \simeq  R_1$ up to small quantum corrections due to irreducible three point correlation functions of two gaugino superfields and $\Phi$.   
 We can  scan the parameter space of superpotential couplings together with $R_1$ (in units of $R_0^{1/3}$)  to see if there are parameter regions where $\delta_{SC}(R_1)  <<1$.   Using   parameters from such regions  of the parameter space  presumably illustrates  the behaviour of the effective theory we may expect: even without  doing the difficult dynamical calculation  of $R_1/R_0^{\frac{4}{3}}$. We show instances  of such parameter regions in an explicit example below (see Table 1.).

\section{A Simple $SU(3)$ based GAM  Example } 
\subsection{GAM with base representation  $r=(3\times 3)_{symm}$ of $SU(3)$} 
As a simple example of condensation in    GAM-AS  systems, consider  a (traceful) GAM $\Phi$    transforming as $R\sim 6\times \bar 6$  of $SU(3)$, i.e. the base representation is $r=6_{SU(3)}$. In this case $d(r)=6,S_2(6) =5/2, d(R)=36, S_2(R)=30, b_0= +21$. Note also that $6\times \bar 6 = 1 + 8 + 27, S_2(27)=27$, so that the $27-$plet irrep alone could be used as the AM-AS system  with $b_0=18$. We here examine  symmetry breaking $G=SU(3)\rightarrow H=SU(2)\times U(1)_Y, Y=Diag(1,1,-2)$  driven by gluino condensation in the UV. The 6-plet of $SU(3)$  which is the   symmetric two (fundamental) index  tensor of SU(3), decomposes under H as \be 6_{\alpha\beta}\equiv 3_{(\bar \alpha \bar \beta)}[2] +2_{3\bar \alpha}[-1] +1_{33}[-4]  \quad ;\qquad \bar \alpha,\bar\beta=1,2 \, (T_3=\pm 1/2, Y=+1)\ee while the  SU(3) triplet index 3 has  $T_3=0, Y=-2$.  SU(2) irreps are identified by dimension and $Y$ quantum numbers are given in square brackets.  Then \be 6\times \bar 6 \equiv  (1 +3 +5 )[0] \oplus  (1+3)[0] \oplus  1[0] \oplus \cdots \ee  The diagonal $3\times 3, 2 \times 2, 1 \times 1$ blocks of $\Phi$ are occupied by  representations that are Y singlets and contain SU(2) singlets so that  \be \langle \Phi \rangle = Diag( V_1\, {\cal{I}}_3, V_2\, {\cal{I}}_2,V_3)\ee where $V_I,I=1..3$ (not all equal if symmetry is to break)  are chosen from $\{v_i^{(q)}; i=1..n=2\}$ and  ${\cal{I}}_m$ is the $m$-dimensional identity matrix. $<\Phi>$  breaks the gauge group    $SU(3)$ to $SU(2) \times U(1)_Y$.  Three  possibilities are $A: V_1=V_2\neq V_3 ;  B:  V_1=V_3\neq V_2 ; C:  V_2=V_3\neq V_1 $ and in each case either the equal pair  or the single VEV can take the value $v_1^{(q)}$. 

Fermion mass terms  from the superpotential  arise   in the pattern $\lambda  \psi_I^J \psi_J^I  (-(v_1^{(q)}+v_2^{(q)})  +(V_I +V_J ) ); I,J=1...6$. Clearly if $V_I,V_J$ are distinct i.e. $V_I+V_J = v_1^{(q)} +v_2^{(q)}$ the mass term will vanish. In  case A this can happen for 5 index pairs(1/2/3/4/5 paired with 6),  Case B  for 8 base rep index pairs (4/5 paired with 1/2/3/6)  and in Case C for 9 (1/2/3 with 4/5/6). Thus these are the numbers of off-diagonal index pairs  (out of the total of 15 ) which remain massless. Since we expect Dirac partners   only for the  4 gauginos of the coset $SU(3)/(SU(2)\times U(1)_Y)$,   pseudo Goldstone(PG)  multiplets  arise. The 4 coset gauginos transform as two doublets of $SU(2)$ with $Y=\pm 3$ and pair up with  2 doublet pairs from the above enumerated massless pairs of conjugate fields. The 6 diagonal pairs  are of course always massive.    For case A $\psi_{\bar\alpha 3}^{33} (\psi_{33}^{\bar\alpha 3})$ teams up with $\lambda_3^{\bar\alpha}(\lambda_{\bar\alpha}^3)$ respectively while $\psi_{\bar\alpha \bar\beta}^{33}$( and $\psi^{\bar\alpha \bar\beta}_{33} $) remain massless. However we shall see that introduction of rank breaking fields gives these putative PG (PPG) multiplets a mass. 
 
 It is perhaps worth emphasising again  that  the basic calculations of $v_i^{(q)}, W^{(q)}$  which define the effective   SBGT of the perturbative GUT    for any other GAM type model with a  degree $n+1$  superpotential  but  different  gauge group, different base representation and   different choices of $\{N_i\}$  are almost identical. It is only the group theoretic analysis of the breaking patterns produced by placement of the derived VEVs on the $\Phi$ diagonals, and of course- should one ever actually dare to calculate them - the derivation of the input dynamical condensate, $R_{0,1}$ etc., that differ in each particular case. 
For example  consider a different GAM say $r_G=10_{SU(5)}$ and  GAM $\Phi \sim 10\times {\overline{10}} = 1+ 24 + 75$, with cubic superpotential.  Then    we only need to reanalyse the allowed placement patterns of $v_{1,2}^{(q)}$ on the diagonal of $\Phi$ and the resultant little groups  while using the different values of $N_1$ allowed by $d(r)=10$. The expressions for the evaluation of the VEVs $v_i^{(q)}$ suffer  only trivial modifications. Thus rather than regarding the pure GAM case as a toy model for a GUT it should be seen as a common structural component for all ASGUTs based on GAMs.

 \subsection{Spontaneous Fractionation of the Gaugino  condensate }
 The physical   gaugino condensate defined by the UV condensed phase is G-invariant and thus cannot discriminate between different gauge equivalent gauginos. However in any given  spontaneously broken phase a distinction between gauginos   associated with the little group($H$)  and  coset ($G/H$)   generators is meaningful. 
The ur-gaugino condensates ($R_{0,1}$ for a cubic  GAM model)  are   physical G-singlet quantities that are   backgrounds relevant  for the computation of quantum VEVs in the spontaneously broken phases of the system  in Regime II. We cannot compute them and thus they are input dimensionful parameters for the   phases decided by the VEVs  of the chiral supermultiplets.  On the other hand the CDSW formalism computes contour integrals of the resolvents $R(z),T(z)$    (that can be solved  for in terms of the input parameters $R_{0,1},\lambda$ etc.) around  the A-cycles of the Riemann surface defined by the quantum dynamics or, in more prosaic terms, around the branch cuts defined by the ``quantum superpotential derivative'' ($y(z)= {\sqrt{ W'(z)^2 + f(z)}}$  in the GAM case).  In the IR strong models the coset ($G/H$) gauginos decouple from the low energy physics and these integrals give the gaugino condensates ${\cal{S}}_i$ corresponding to the different $U(N_i)\in H$ factors. 

{\emph{Now in the AS case the situation is radically different}}. All the gauginos are still in play in the full spontaneously broken gauge theory of Regime II i.e. at and just below  Grand Unified scales $\sim M_X$. However the SSB pattern coded in the $\{N_i\}$ partition of $d(r)$ assumed as dynamically invariant input implies that the integrals of the resolvent $R(z)$  will now yield the condensates of  certain characteristic combinations of gaugino bilinears since the contour integral  $\oint_{A_i} dz \,Tr (T^A_{r} T^B_{r} (z-\Phi)^{-1})$   will give a non-zero result only for the  $i-$th VEV/branch cut. This implies that only certain combinations of the gauginos will contribute to the integral computed using the solution for $R(z)$. The peculiar indirect manner in which the little group is specified by the placement of $v_i^{(q)}$ on the diagonal of $<\Phi>$   determines the combinations. Thus for example in Case A  where $ \Phi =Diag(V_1{\cal{I}}_5,V_3)$ the relevant combinations are  determined by the partial traces  defined by including a projector on to the subspace corresponding to the cycle $A_i$ : 
\bea W_\alpha^A W^{B\alpha} Tr T^A_{(r)} T^B_{(r)} Diag({\cal{I}}_5,0)=  {\frac{5}{2}}\, {\vec W}^2 + 2 \sum_{\bar A=4}^7  {W_{\bar A}}^2 + {\frac{7}{6}} \, {  W_8}^2  \nnu
W_\alpha^A W^{B\alpha} Tr T^A_{(r)} T^B_{(r)} Diag({\slashed{0}}_5,1)= 0\, {\vec W}^2 +  {\frac{1}{2}} \sum_{\bar A=4}^7  {W_{\bar A}}^2 + {\frac{4}{3}} \, {  W_8}^2  \eea

The gaugino sub-condensate  patterns  are described by the  values of   the  SU(2) triplet condensate  $\kappa   {\vec W}^2 $, $SU(3)/(SU(2)\times U(1)_Y)$ coset condensate $ \kappa \sum_{\bar A=4}^7   W_{\bar A} ^2 $   and the  hypercharge gaugino  condensate   $ \kappa {  W_8}^2 $.  The total of the coefficients of each gaugino squared  is equal to  $S_2(6_{SU(3)})=5/2$ as expected, confirming  that the division corresponds to a fractionation of the gauge singlet condensate following the SSB pattern of the phase in question. The computation of the trace can be easily carried out by setting up the $6\times 6$ generators of $SU(3)$ in the 6-plet representation using the symmetric 6-plet generators obtained from the symmetrized tensor product:  ${T^{(6)A}}_{(ij)(kl)}=(1/2) ( {\delta}_{l(j}{T^{(3)A}}_{i)k} +\delta_{k(j} {T^{(3)  A}}_{i)l}  )$. These combinations are  
  denoted compactly  by giving the coefficients for each of these VEVs : In Case A (5/2,2,7/6) for $V_1=V_3$ and (0,1/2,4/3) for $V_2$.  Case B :  coefficients (2,5/4,7/3) for $V_1=V_3$ and (1/2,5/4,1/6) for $V_2$ and in Case C  (1/2,7/4,3/2) for $V_2=V _3$ and (2,3/4,1) for $V_1$.     
  
  \subsection{Numerical investigation of semi-classicality}
In Table 1 we give instances of  the calculation of the vacuum expectation values for this model taking $m=\mu=0$, for simplicity,  for which  $\delta_{SC} $ defined in eqn(\ref{deltaSC}) is indeed small  and the semiclassical approximation $R_1\simeq \sum_j R_{0j }v_j$ is good. Note that the nonzero values of $v^{(q)}_i$ obtained are in units of $R_0^{1/3}$ and there are no nonzero  VEVs at the classical level. It is convenient to rescale to dimensionless (hatted)  variables in units of $R_0^{1/3}$  thus : $z=\hat z R_0^{1/3}, \, f_0 = \hat f_0 R^{4/3}$  and so on. It is easy to find  values of the free parameter $(\lambda,\hat R_1$)  sets for which accurate semi-classicality(i.e. $\delta_{SC}<<1$) can be attained. However a numerical calculation of the condensate values actually obtained in the strong coupling region is beyond our abilities at this stage.

\begin{table}
\begin{center}
\begin{tabular}{|p{.4cm} |p{.6 cm}|p{1.4cm} |p{1.4cm}|p{1.4cm} |p{1.4cm}|p{1.4cm} |p{1.4cm}|p{1.4cm} |p{1.4cm}|p{.4cm}}
\hline
  $N_1 $ &$d(r)$ &$\lambda$&$\hat R_1$&$\hat t_1$ &$\hat v_1 $&$\hat v_2$ & $\hat R_0^{(1)} $&$\hat R_0^{(2)} $&$\delta_{SC}$\\
 \hline
     $4 $ &$6$ &$ -1.582    -.295 i$&$ .247+.374 i $&$  -.0328     +.443 i$ &$-.220    +.630 i$&$ .424     -1.038 i$ & $.772    -.192 i$&$.228     +.192 i$&$0.000$\\  
  \hline
  $4 $ &$6$ &$1.381  -.0680 i$&$-1.001   +.487 i$&$-1.322   +.241 i$ &$-.782 $&$.904   +.135 i$ & $1.124   -.299 i$&$-.124   +.299 i $&$.0009$\\  
 \hline
    $2 $ &$6$ &$0.4$&$0.595    + 0.995i$&$-3.540   +   4.186i$ &$  0.201    + 2.72 i$&$-.986   - .312i$ & $0.566   - 0.293i$&$.434    + .293i$&$.003$\\ 
   \hline
    $2 $ &$6$ &$0.4$&$0.595 - 1.005i$&$-3.489   -4.169 i$ &$0.180   -2.704i $&$-0.962   + 0.310i$ & $0.566   +0.299i $&$0.434   -0.299i $&$.002$\\  
   \hline
    \end{tabular}
\caption{Illustrative values obtained by searching for values of $\hat R_1$ that minimize $\delta_{SC}$ given $N_1,\lambda $  for the classically mass scale free model with $d(r)=6$ and symmetry breaking $SU(3)\rightarrow SU(2)\times U(1)_Y$. All dimensionful quantities are in units of $R_0^{1/3}= 5^{1/3}\Lambda$  and the vacuum expectation values are due purely to quantum effects. The occurrence of such regions of parameter space supports the possibility  of solutions where the quantum corrections driven by the gluino condensate are captured  by the quantum VEVs  in the sense that $R_1\simeq \sum R_0^{(j)} v_j $.}  
\end{center}
\end{table}

 \section{Rank Reduction }
 \subsection{$r$ based rank reduction}
Since the elements of $\Phi\sim r\otimes \bar r $ with VEVs are neutral w.r.t. all  the Cartan subalgebra  generators  the gauge group rank cannot decrease  due to  symmetry breaking in any purely  AM type model.  However    GUT models (such as those based on  $SO(10)$ ) with rank $\geq 5$  require rank reduction  to break the gauge symmetry to the SM gauge group which has rank 4. In the Minimal SO(10) GUT \cite{aulmoh,msgut}, just such a rank reduction is achieved by including a pair of conjugate representations ($126,{\overline{126}}$) whose role is precisely to break $SO(10)\rightarrow SU(5)$.    The authors of \cite{GKA} have also provided \cite{seiberg2,GKA2} an analysis for  Adjoint-SYM with Flavours,    i.e. super $SU(N_c) $ YM with an adjoint as well as  $N_f$   pairs of Quark  and anti-Quark  chiral multiplets     $Q_f,\bar Q^f, f=1...N_f$.  Such models  possess ``Higgs vacua"   with $<Q_f>,<\bar Q_f> \neq 0, f=1..,n $, which imply    rank reduction $  N_c -1 \rightarrow N_c-n-1 $ which is  unachievable with AM type fields alone.
 
 Consider  $N_f=1$ i.e.  with a  pair of  complex representations $Q,\bar Q$ transforming as $r,\bar r$  added to GAM $\Phi\sim r\times \bar r$. For this simple extension the extra Chiral ring generators  $\bar Q \Phi^n Q, n=0,1,2....$   may all be solved for.  On the other hand by defining
 an GAM  type composite  field $\chi \equiv  Q {\bar Q} $  and considering $ Tr \Pi_{n_i } (\chi  \Phi^{n_i}) $ as well as similar expressions with   an insertion of $W_\alpha W^\alpha$ it is clear that one can write down a much larger set of Chiral operators  than we can easily solve for.  The size of the Chiral ring increases rapidly with  each new chiral multiplet introduced to the model and the GKA procedures do not permit solution for the full set of Chiral operator VEVs.  Nevertheless we shall see that one can still extract the information required to define the effective perturbative SB GUT generated by dynamical symmetry breaking driven by gaugino condensation at the  UV  Landau pole. 
 
     We modify the superpotential while maintaining  renormalizability by adding gauge invariant terms ($\Delta { {W}}= { {W}}_Q= - \eta \bar  Q\cdot  \Phi  \cdot  Q$).  We have omitted a  Quark mass term by shifting $\Phi$. The model then admits semi-classical ``Higgs vacua''  in which parallel  components of the complex multiplets $Q,\bar Q$, say $Q_1,\bar Q_1$, obtain VEVs (of equal magnitude to cancel the D term contributions)  leading to a lowering of the rank of the little group by one :
 \bea 
 <\Phi_{11}>=z_1&=&0   \qquad  ; \qquad <Q_1>= \sigma,\quad\, <\bar Q_1>= \bar \sigma   \nnu
 |\sigma|&=&|\bar \sigma | \qquad ; \qquad    \sigma  \bar \sigma = \frac{W'(0)}{\eta} \eea

Since $W_\alpha^A (T^A)_{ij} Q_j=    W_\alpha^A (T^A)_{ij} \bar Q_i  \simeq 0  $ in the Chiral ring,  the  loop equation containing $R^2(z)$ is unchanged from the pure AM case(eqns(\ref{ResR},\ref{Rqdef}))    but the equation for $T(z)$  is modified to :
\bea &&(2 R(z) -W'(z)) T(z) +\eta S(z) -{\frac{c(z)}{4}}=0 \label{Tloop} \nnu
&& S(z)\equiv  \bar Q \cdot {\frac{1}{z-\Phi} } \cdot Q  \equiv \sum_{n=0}^{\infty} {\frac{S_n}{ z^{n+1} }}     \eea 
  The GKA for  $\delta Q \equiv  {\frac{1}{z-\Phi} }  \cdot  Q $  or $     \delta \bar  Q\equiv \bar Q \cdot   {\frac{1}{z-\Phi} }  $  give  
\bea S(z) &=& \frac{-R(z)  +\eta  \bar Q\cdot Q}{\eta \, z  }\eea
    Then if we impose \bea {\frac{1}{2 \pi i}} \oint_{C_{z_1}} T(z) dz =1 \eea   we find 
\bea  {\bar Q}_1  Q_1 &=& {\frac{R(0)}{2 \eta}} =   {\frac{W'(0) +y(0) }{2 \eta }} \label{Q1Q1barsol}\eea 
 Which reverts to its classical value $W'(0) / \eta$  as $y(z) \rightarrow W'(z)$ i.e.  in the absence of quantum effects a.k.a gluino condensate.  The VEVs of $Q,\bar Q $ lead to useful mass matrix contributions. For example for a SU(3) model based on the 6-plet (i.e $\phi\sim 6 \otimes \bar 6$)  discussed above we take $Q_6=Q_{(33)}=\sigma,\bar Q_6=\bar Q^{(33)}=\bar \sigma $. These VEVs Dirac-pair $Q_{(33)}, Q_{(\alpha 3)} $with gauginos $\lambda_3^{ 3},\lambda _3^{ \alpha} $ respectively. Moreover they modify the masses of the putative pseudo-Goldstone (PPG)  multiplets.  For example in case A where $V_{(\bar\alpha\bar\beta)}=V_{\bar\alpha 3}\neq V_{(33)}$, the PPGs    $\Phi_{(\bar\alpha\bar\beta)}^{(33)}, \Phi^{(\bar\alpha\bar\beta)}_{(33)}  $  Dirac-pair  with $\bar Q^{(\bar\alpha\bar\beta)}, Q_{(\bar\alpha\bar\beta)} $   and become massive.  
       
       We can define an effective cubic  superpotential that works to reproduce the quantum VEVs along the same lines as in the pure adjoint case by modifying the parameters of the cubic superpotential   so as to support a Higgs vacuum solution with $\Phi_{11}=0$ even in the quantum case but with:
       \bea  \lambda^{(q)}&=&\lambda \qquad;\qquad m^{(q)}= - 2 \lambda (v_+^{(q)} +v_-^{(q)} )    \nnu
       \mu^2_{(q)}&=&     \lambda  v_+^{(q)}  v_-^{(q)}    \qquad;\qquad  \eta^{(q)} ={\frac{2\,   \mu^2_{(q)}}{W'(0) + y(0)}} {\, \eta } \eea 
       which reproduces $v^{(q)}_\pm$ as possible solutions for $\Phi_{nn} , n\geq 2$ but also  eqn(\ref{Q1Q1barsol}) for $<Q_1,{\bar Q_1}>$.

              \subsection{ $r \times r$ based rank reduction}
                  Besides   rank reduction based on complex representations $\sim r,\bar r$, one can also consider more complicated scenarios based upon pairs of symmetric  representations $\sim (r\times r)_s,(\bar r\times \bar r)_s  $  which prove useful in realistic scenarios(see the next section).   Following and extending \cite{naculich,eynardZJ},  we  survey  the solution of the resolvent system for this case. Introduce a pair  of chiral supermultiplets  $\Sig_{ij}=\Sig_{ji},\Sigb_{ij}=\Sigb_{ji}; i,j=1...d(r)$ transforming as $(r\times r )_{symm}, (\bar r\times \bar r )_{symm} $  and an additional superpotential 
   \bea { {W}}_{\Sig} =  -\eta \Sigb^{ij}  \, \Phi_j^{\, k} \,\Sig_{k i}   \eea 
  As in the $Q\bar Q \Phi$ case there is a semiclassical Higgs vacuum  where one  conjugate component pair  from $\Sig,\Sigb$   (say $(\Sig_{MM},\Sigb^{MM})$)  gets a VEV   :
 \bea  {\Sig}_{MM} &=&\sigma \quad ; \qquad \Sigb^{MM} =\bar\sigma  \qquad ;\qquad   |\sigma|=|\bar\sigma |\nnu
 \Phi_M^{\, M}  &\equiv & z_\sigma  =  0    \qquad ; \qquad   \sigma\,  \bar \sigma= {\frac{ { {W}} '(0) }{\eta}}\eea
  
For example in the case of the SU(3) model based on the 6-plet  $S_{ij}=S_{ji}$, $\Sig,\Sigb$ are $6\times 6$ symmetric matrices and  we can break $SU(3)\rightarrow SU(2) $ by 
\bea  \Phi   &= &Diag(V_1,V_1,V_1,V_2,V_2,0)  \qquad ; \qquad   V_i \in \{ v_+^{(q)},v_-^{(q)} \} \nnu
 \Sig &=& Diag(0,0,0,0,0,\sigma)  \qquad ; \qquad   \Sigb   = Diag(0,0,0,0,0,\bar \sigma) \nnu
{\underline{6}}&=& \{(11),(22),(12),(13),(23),(33)\} \eea
 Since $S_{(33)}$ is an SU(2) singlet but has $Y=-4$ it is clear that  the VEVs of $\sigma,\bar\sigma \neq 0$ reduce the rank by 1. As in the case with $d(r)$-plet rank-breaker pairs $\bar Q,Q$ the PPG spectrum becomes massive due to the rank breaking VEVs. 

By considering   a combination of the loop equations for GKA variations \bea \delta \Phi &=& \kappa W_\alpha W^\alpha (z\pm \Phi)^{-1} \quad ;\quad \delta \Sig = 2 \kappa W_\alpha (z-\Phi)^{-1} \cdot  \Sig \cdot  (W^\alpha (z+\Phi)^{-1})^T \eea and using the Chiral ring constraints ${W^\alpha}_{(i}^j \Sig_{k)j}  =0 $ (and similarly for $\Sigb$)  one derives\cite{naculich}  the loop equation  (notation $\bar F(z)\equiv   F(-z)$)  \bea R(z)^2 +\bar R(z)^2 +R(z) \bar R(z) -W'(z) R(z) -{\overline W}'(z) \bar R(z) = r_1(z)\equiv {\frac{f(z) +\bar f(z)}{4}}\label{eqRRb}\eea   For  a cubic superpotential $r_1= f_0/2= -2 \kappa Tr (\lambda \,  \Phi +m) W_\alpha W^\alpha  $. Due to  branch  cuts   in the $z$ plane (that emerge  further on)  the resolvent function $R(z)$ is {\emph{not}}  even in $z$. Introducing a new resolvent  (the analogue  was automatically zero in the previous case due to  the Chiral ring constraint $W_\alpha \cdot Q \simeq 0$)  :
\bea U(z) \equiv \kappa \Sigb \cdot {\frac{ W_\alpha W^\alpha }{z-\Phi} }\cdot \Sig  \label{Ueqn}\eea 
we obtain a  modified equation for $R^2$ : \bea R^2 -W'(z) R = {\frac{f(z)}{4} } -\eta\,  U(z)  \eea and a similar equation for $\bar R$.   One can then show  \cite{eynardZJ,naculich} that  by substituting  $R(z)=\omega(z) +\omega_r(z),\bar R(z) =\omega(-z) +\omega_r(-z)\,\, ; \, \omega_r(z)\equiv (2 W'(z)-{\overline W}'(z))/3$,  eqn(\ref{eqRRb}) simplifies  to just $\omega^2 +\bar\omega^2 +\omega \bar \omega= r(z)=r_0(z) +r_1(z) \,\, ; \, r_0\equiv (W'^2+{\overline W'}^2+W' {\overline W'} )/3$.  Then  we obtain  $\omega=u_1(z),\bar\omega =u_3(z), -(\omega +\bar\omega)=u_2(z)$ where    $u_a,a=1,2,3$ are solutions of a cubic equation of a special form 
\bea u^3(z)  -r(z) u(z) -s(z)  = \prod_{a=1}^{a=3} (u-u_a(z) )=0\eea Here 
$ s(z) = s_0(z)  +s_1(z) $  and $s_0,s_1$ are polynomials of degree $3n,2 n-2$ which can be  explicitly calculated \cite{naculich} given $W(z)$ :
\bea s_0(z) &=& \omega_r \bar  \omega_r (\omega_r +\bar \omega_r)  ={\frac{2}{27}}(\lambda  z^2 +\mu^2) ((\lambda  z^2 +\mu^2)^2 -9 \, m^2 z^2)\nnu
s_1(z) &=& 2 \eta (m\, U_0  +\lambda\, U_1)  +{\frac{1}{4}}(\omega _r \bar f(z)  +\bar  \omega _r  f(z) )\nnu 
&& = 2 \eta (m\, U_0  +\lambda\, U_1)      -{\frac{2}{3}} (m\,  R_0 (\mu^2 - 2 \lambda z^2 ) + 
   R_1 \lambda  (\mu^2 +  \lambda  z^2  ))\eea
Thus one finds 
\bea \omega(z) = e^{-{\frac{2 \pi i}{3}}} \omega_++e^{+{\frac {2 \pi i}{3}} }{\frac{r(z)}{3 \omega_+}}  \qquad ; \qquad \omega_+=   \big ({\frac{s(z)}{2}}  +\sqrt{ {\frac{s^2}{4}}  -{\frac{r(z)^3}{27}}}\big )^{\frac{1}{3}}\eea
In spite of the cube root,  the Riemann surface  branching structure for $\omega(z),\bar \omega(z)$ is still two sheeted provided  $\omega_-(z)=\omega_+(-z)= {\frac{r(z)}{3 \omega_+}} $. The third root $\omega(z) +\bar\omega(z)$ occupies an isolated `singleton sheet'.  Since $s(z),r(z)$ are even polynomials the condition on $\omega_\pm$ can be satisfied provided  $ \sqrt \Delta\equiv  \sqrt{{\frac{s^2}{4}} -{\frac{r^3}{27}}  }$ is an odd function. This can be ensured by imposing a constraint fixing a higher $R_n$ coefficient in terms of a lower $R_n$ coefficient.  Writing $\Delta(z)= z^2 Q(z^2)=z^2 P(z)$, one finds  that  for non-zero $m,\mu^2$ the polynomial $P(z)$ defining the  branch cuts and Riemann surface is quartic in $z^2$ i.e. even and of degree 8 in $z$ and has 4 square-root branch cuts defining  a Riemann surface of genus 3. Contour integrals around these branch cuts  play the same role as in the pure  AM case. {\emph{Thus we expect  4 possible quantum VEVs }}  when the tree level superpotential is cubic,  even though at tree level there are just two semi-classical VEVs $v_\pm \neq 0$ besides the vanishing VEV of $\Phi_{11}$.  This indicates that the effective superpotential in the general case with $m,\mu^2\neq 0$ will need to be quintic in $\Phi$.  

Since the analysis becomes quite involved for the general case we here present the  explicit solution of the resolvent system only for $m=\mu^2=0$. In some sense this solution is more interesting  since it eliminates  all explicit  mass scales completely so that all masses arise purely by dimensional transmutation and the classical theory will be superconformal. Moreover, since $\Delta(z)$ is sextic and $P(z)$ is quartic,  there are only two branch cuts and thus two quantum VEVs. A  cubic quantum effective super-potential can still be defined  as in the earlier cases studied. 

  We now trace the determination of resolvent coefficients $\{U,R,T,S\}_n$ using the available GKA equations. Firstly the expansion of eqn(\ref{Ueqn}) and then eqn(\ref{eqRRb}) for large $z$ determines $U_n,R_{2 n+3}, n=0,1,2....$ in terms of $R_{2n}, R_1$  : 
\bea U_0 &=&  { \frac{\lambda R_2}{\eta}}  \quad ; \quad U_1= -{ \frac{  R_0^2}{2 \eta}} \nnu
 U_2&=& { \frac{ \lambda R_4 -2 R_0 R_1 }{\eta }} \quad ; \quad U_3=  { \frac{   R_1^2 -2 R_2 R_0}{ 2 \eta}} \nnu  R_3 &=&{\frac{R_0^2}{ 2 \lambda}}\quad\qquad \qquad;\qquad R_5 = {\frac{(3 R_1^2 + 2 R_0 R_2)}{ 2 \lambda}}.... \eea

Next imposing $\Delta(0)=0$ fixes $R_1$  and with   $\Delta(z)=z^2 P_4(z)  $ we have ($s_{1,2}=\pm 1$)
\bea R_1 &=&  (-{\frac{27}{32 \lambda } } R_0^4)^{\frac{1}{3}}\nnu
  P_4(z) &= &-{\frac{1}{432}} R_0^2 \lambda^{ \frac{8}{3} }(108   (2 R_0^4)^{\frac{1}{3}} +9   (2  \lambda R_0)^{\frac{2}{3}} z^2 +16 z^4 \lambda^{\frac{4}{3}} )\\
&=&- {\frac{\lambda^4 R_0^2}{27}}  \prod_{s=\pm} (z-z^{(+s)})(z-z^{(-s)}) \quad;\quad  
z^{(s_1, s_2)} =  {\frac{s_1}{4}} ({\frac{R_0}{{\sqrt{2} \, \lambda}} )^ {\frac{1}{3}}}     \sqrt{-9 +  s_2   15\,  i \sqrt{15}} 
 \nonumber\eea
  Thus the two branch cuts in this (degenerate) case run between $z^{++},z^{+-}$ and  $z^{-+},z^{--}$.  We emphasize that   the determination of $R_1$ in terms of $R_0$  is a novel consequence of adding $\Sig,\Sigb$.  This simplifies  the numerical  analysis  significantly since-after rescaling to dimensionless  form- only the dimensionless coupling $\lambda$ remains free. Contrast this with the pure AM case where $R_1$ was to be dynamically determined.

Now we can also obtain all the even coefficients $R_{2n}, n=1,2,...$ by expanding the cubic equation for $\omega(z)=R(z)-\omega_r(z)$ for large $z$.  This gives 
\bea R_2=9 ({\frac{R_0^5}{2^{13}\lambda^2}})^{\frac{1}{3}} \quad ;\quad R_4  =-{\frac{1233}{512}} ({\frac{R_0^7}{4\lambda^4}})^{\frac{1}{3}} \qquad ; ....\eea

Finally  we  define the pure rank-breaker resolvent   $ S(z)\equiv  \Sigb \cdot  (z-\Phi)^{-1} \cdot \Sig $.  As in the $\Phi Q  \bar Q $ case one derives the system of GKA resolvents 
\bea   
\eta S(z)    &=&    {\frac{c(z)}{4}}   +({ {W}}'(z) -2 R(z))T(z)               {\label{eqS}}\eea
where $f(z),c(z)$ are as before. Thus given $T(z)$   one can derive  $S(z)$. To find $T(z)$ we use\cite{naculich}  the equation   which is the analogue of eqn(\ref{eqRRb})  derived using the same $\Phi$ variations  but with  $\kappa W_\alpha W^\alpha$ factors omitted :
\bea   {\frac{c(z) +\bar c(z)}{4}} =(2 R(z)  -W'(z)) T(z)  +(2 \overline R(z) -{\overline W}'(z)) \bar T(z) \nnu +(R(z) \bar T(z) +\bar R(z)  T(z))  +2 {\frac{\bar R(z) -R(z)}{z}} \label{eqTTb}\eea 
Motivated by the solution of the   $\bar Q Q \Phi$ case  where the corresponding equation differs only by the absence, on the r.h.s.,  of the mixing (third) term and the factor of 2 in the fourth term  and has solution  $T(z)=(2 R-W'(z))^{-1}((R-\eta \bar Q Q)/z   +c(z)/4)$ we propose 
\bea T(z) = {\frac{1}{(2 R -W'(z) +\bar R)} }({\frac{c(z)}{4}}+{\frac{ 2 R(z)}{z}} +\zeta(z))\label{solT}\eea
where $\zeta(z)$ is to begin with an arbitrary {\emph{odd}} function of $z$. However the behaviour of $T(z)$ as $z\rightarrow \infty$ allows only $\zeta(z)=-\lambda T_2 /z$.  Here $c(z) =  -4 \lambda( T_1 + T_0 z)$.     By expanding   eqn(\ref{solT}) for large $z$ we get   $T_{n>3}$. If, following the pattern of the Higgs vacuum solution in the $\bar Q Q \Phi$ case we demand that   the  residue of $T(z)$  at $z=0$ be unity, corresponding to the rank breaking Higgs vacuum,  we determine $T_2$ :
\bea   T_2 = (-{\frac{ R_0^2} {  2 \lambda^2 }})^{\frac {1}{3}} \quad ; \quad T_3=  {\frac{(T_0-2) R_0}{\lambda}} \quad ;\quad  T_4=  {\frac{ R_0 T_1}{\lambda}} - 3 (3 T_0-2)  ( {\frac{ R_0^4}{32 \lambda^4}} )^{\frac{1}{3}}.....\eea
  Just as in the  $\bar Q Q \Phi$  case eqn(\ref{solT})  gives the correct $T(z)$ in the semiclassical limit where $R,\bar R\rightarrow 0 $. Of course the semi-classical limit is trivial in this massless  case in the sense that all VEVs are then  zero.   The quantum superpotential derivative   is now   
 \bea y(z)\equiv W'(z)- 2 R(z)-\bar R(z)    = -(2\, e^{-{\frac{2 i \pi}{3}}} +e^{\frac{2 i \pi}{3}})\omega_+(z) -(2 \, e^{\frac{2 i \pi}{3}} +e^{- \frac{2 i \pi}{3}})\omega_-(z)  \eea 
 Note that the square root branching structure of $y$ is now hidden inside the expressions for $\omega_\pm$ which contain ${\sqrt {P(z)}}$.  Where $P(z)$ is the quartic($m=\mu=0$)/octic ($m,\mu$ non-zero) polynomial which defines the branch cuts. The resolvent for $S(z)$ is also determined via eqn(\ref{eqS}) once $R(z),T(z)$  are  known so that the coefficients $S_n$ of $z^{-n-1}$ in the large $z$ expansion can be read off. Thus  we get 
  \bea  \eta  S_0 &=&  \lambda T_2 =  (-{\frac{ R_0^2 \lambda }{ 2 }})^{\frac{1}{3}} \quad ; \qquad \eta S_1  =  -R_0 (T_0 +2) \nnu
  \eta S_2 &=&  R_0 T_1 - 3(T_0-2) ({\frac{R_0^4}{32 \lambda}})^{\frac{1}{3}} \eea
  and so on. The residue of $\eta S(z)$ at  $z=0$ is  $(-4  \lambda R_0^2  )^{1/3}$. 
  
  Since $T_0=d(r)$  we have only $R_0,T_1$ left undetermined. The former is set by the gaugino condensation ${\cal{S}}=\Lambda^3$ .  The same argument as for the case with the Adjoint gives 
   \bea T_1 &= & {\frac{  2 \pi i N_1-\oint  dz \, (\lambda \, d(r) z  + (\lambda T_2 -2 R(z)) /z ) y^{-1} }   {\oint  dz\,\lambda \,  y^{-1}} } \eea
where $ y(z)$ was given explicitly above. Thus with input parameters $\lambda,  N_1$ ($N_2\equiv d(r)-N_1-1$)
and using units of   $R_0^{\frac{1}{3}} \sim \Lambda $ for dimensionful quantities we can evaluate the quantum VEVs by performing the contour integrals numerically.  Details will be given in a sequel.

Although we also  defer detailed  consideration   of the case with $m,\mu\neq 0$  to the sequel it is important to underline  that it presents new features not observed in the degenerate  case described above. One finds $\sqrt{\Delta} = z \,\sqrt{ Q_4(z^2)}= z \sqrt{P_8(z)}$ so that one has 4 rather than 2 branch cuts. This opens the possibility of cases where the quantum spontaneous symmetry breaking includes VEV patterns with no semi-classical antecedent. With cubic $W(z)$  and thus   two critical points, the contour integrals around the 4 branch cuts    define 4 different VEVs : which may be placed at will on the diagonal of the quantum corrected VEV of $\Phi$  giving a symmetry breaking pattern without a semi-classical analog.   Thus the corresponding ``perturbative effective quantum superpotential '' will need to be quintic  rather than cubic. 

By solving the Loop equations we have determined most elements of  the infinite sets of Chiral VEVS $\{R_n, T_n, S_n, U_n\}$.  This  is sufficient for the purpose of defining the effective perturbative  SYMH theories  with gauge symmetries broken by VEVs driven by gaugino condensation and dimensional transmutation. However, as remarked earlier for the case of Quark type additions,  one can define a further vast set of Chiral invariants by defining $\chi_i^j \equiv \Sig_{ik} \Sigb^{kj} $ and then considering $ Tr \Pi_{n_i,m_i}  \chi^{n_i} \Phi^{m_i} $ as well as similar expressions with   an insertion of $W_\alpha W^\alpha$.   The determination of these VEVs is beyond the scope of this paper.

  \section{Realistic MSGUT type model}
  We next come full circle and  consider  our motivating problem: the gauge UV   Landau pole in the successful MSGUT\cite{msgut}. To illustrate how the AS dynamics permits novel realistic GUT scenarios with dimensional transmutation and dynamical symmetry breaking,  we propose  a realistic Spin(10) gauge model with  3 matter $16$-plets   and a Higgs structure generated by  base representation $r=16$ . We take  $\Phi\sim 16\times{\overline{16}}=1+45+210,\Sig  \sim 16\times 16 =10+120+126,  \Sigb  =  {\overline{16}}\times{\overline{16}}=10+120 +{\overline{126}}$.  Thus the sub-irreps present cover all the  irreps used in MSGUT type models \cite{aulmoh,msgut}.  As before we note that one may choose to work with just the irreps of the MSGUT, or some extended set thereof, by applying projectors to select only the irreps one wishes to keep. In the case at hand one specifies $\Phi\sim  210$  by imposing $Tr \Phi=Tr T^A_{(16)}=0$   and similarly on its variations when deriving the GKA relations. Similarly we can extract the 10,120,126-plet  from $(16\times 16)_s$ and ${10,120,\overline{126}}$-plet   from $({\overline{16}}\times {\overline{16}})_s$ using the  Clifford algebra matrices for Spin(10)\cite{alaps}.   This implies   a great increase in the number and types of  resolvents that one must consider and the methods for handling them do not yet exist.  Thus we stick with the full tensor product reps, specially since this entails no cost in terms of additional parameters. 
  
  The superpotential for the complete model as 
  \bea W={\frac{m}{2}} \,Tr \Phi^2 + {\frac{\lambda}{3}}\, Tr \Phi^3   + \mu^2 \,Tr \Phi  -\eta\, \Sigb \cdot \Phi \cdot \Sig \nnu +h_{AB}\,\Psi_A \cdot  \Sigb \cdot \Psi_B  +h'_{AB} \, \Psi_A \cdot  \Sig  \cdot \Psi_B {\label{Wspin10}} \eea
  where $\Psi_A,A=1,2,3$ are the three matter 16-plets and  it is understood that in the last term  only the real 10-plet and 120-plet parts of the tensor product will be present since there is no invariant between two matter 16 plets ($\Psi_A $)and a 126-plet.  Notice the remarkable economy of AM type couplings.  We   recall our proposal\cite{yumguts} to further reduce   the number of matter Yukawa couplings by making $\Sig,\Sigb$ carry the generation indices.  The analysis of the SSB associated with the first line of eqn(\ref{Wspin10}) closely follows the  method for GAM in Section 2  and symmetric tensor product based  rank reduction in Section 4  and the case with $m=\mu=0$ is equally interesting for the realistic MSGUT type model. The common skeleton of the EV extraction process has already been described. Thus we need to indicate only   the group theoretic  features particular to the determination of possible little groups for this model.
  
  The   Yukawa coupling of $\Sig,\Sigb$ in the second line of eqn(\ref{Wspin10})  presents distinct new features. 16-plet VEVs break SM symmetry except for right handed sneutrino VEVs and those still violate R-parity destroying one of the most signal and  phenomenologically desirable  features of MSGUTs\cite{msgut}. Thus we shall assume that the matter 16-plets have no VEVs.   However  the Konishi anomaly will also  force the development of large purely  quantum trilinear  condensates   involving $\Psi_A \cdot  \Sigb \cdot \Psi_B, \Psi_A \cdot  \Sig  \cdot \Psi_B $, even assuming the  solution has no 16-plet  VEVs.  The theoretical and phenomenological implications of such condensates are not clear to us. 
  
   Moreover the presence of 16-plets further expands the already vast set of Chiral invariants that can be formed from $\Phi,\Sig,\Sigb$ by forming composite 10,120 and 126 plets from bilinears of spinors and then contracting powers of these with powers of $\Phi,\Sig, \Sigb, \Sig \cdot \Sigb $ and so on. We will not attempt to enumerate these new invariants and only note that, for our immediate purpose of characterising dynamical GUT breaking VEVs, the  solution for the VEVs of the full chiral Ring is not required. Thus the complexity of the full Chiral Ring is {\emph{not}} an obstruction to working with this model.

   For analysing the SO(10) VEVs  decompositions w.r.t.  the maximal sub-group $G_{PS}\in SO(10)$  are explicitly available in \cite{alaps} and prove very useful. 
We use  the conventions and results of \cite{alaps} to explicitly calculate the decomposition of Spin(10) invariants. Here   $\mu =\bar\mu,4; \bar\mu =1,2,3  $  refer to the SU(4) indices of the Pati-Salam maximal subgroup $G_{422} =SU(4)\times SU(2)_L\times SU(2)_R  \subset SO(10)$.  Barred mid-Greek indices ($\bar \mu$ etc.) are  colour indices.The fundamental doublet  indices of $ SU(2)_L(SU(2)_R)$ are referred to as $\alpha,\beta(\dot\alpha,\dot\beta)=1,2(\dot 1,\dot 2)$.   It is convenient to order the elements of the 16-plet according to their SM quantum numbers (denoted compactly by the relevant MSSM left-chiral fermion symbol) as 
 \bea 16 &=&\psi_{\mu,\alpha}(4,2,1) \oplus \psi^\mu_{\dot\alpha} (\bar 4,1,2)  =  \{\nu^c[1,1,-1/2,1](4^*,\dot 2),e^c[1,1, 1/2,1](4^*,\dot 1), \\ \,\,\,\, &&   u^c[\bar 3,1,-1/2,-{\frac{1}{3}}]({\bar \mu}^*,\dot 2),  d^c [\bar 3,1, 1/2,-{\frac{1}{3}}]({\bar \mu}^*,\dot 1) \}_L   \nnu &\oplus&\{ L[1,2,0,-1] ({4},\alpha),Q[3,1,0,{\frac{1}{3}}]({\bar \mu},\alpha)\}_L \nonumber {\label{16labels}}\eea
 where we have also  given the dimensions/quantum numbers w.r.t $[SU(3), SU(2)_L,T_{3R}, B-L]$ and the $G_{422}$ indices of each left-chiral matter field.  
Apart from  matter 16-plet VEVs, which must vanish in viable vacua,   the  VEV patterns which can develop will follow the  earlier discussion  in Sections 2. and 4 of the generic form of VEV generation for $\Phi,\Sig,\Sigb$.  The features particular to the model in hand are just group theoretic tracing of the pattern of SSB given the  quantum VEVs that can result from the cubic superpotential. As already noted,in the massless case, there are only two non-zero VEVs even when $\Sig,\Sigb$ are present. 
The labelling of the diagonal elements of $\Phi=16\times {\overline{16}}$ that follows from eqn(\ref{16labels}) above is :  
  \be \Phi =Diag(V_{4^*\dot 2} =V_1, V_{4^*\dot 1} =V_2, V_{\bar \mu^*\dot 2} =V_3  \,{\cal{I}}_3,  V_{\bar \mu^*\dot 1} =V_4\,{\cal{I}}_3, V_{ 4\alpha} =V_5\,{\cal{I}}_2,  V_{ \bar\mu\alpha}=V_6\,{\cal{I}}_6)\ee  with           
 $ \Phi_{11} \equiv \Phi_{\nu^c { \nu}^{c*}}= V_1=0 $ as per the 16-plet  labels introduced above. Then the rank breaking VEVs will  be  $  \Sig  =Diag(\sigma, {\slashed{0}}_{15})\,$  i.e. $\Sig_{4^*\dot 2,4^*\dot 2} =\sigma $,  $\Sigb=Diag(\bar\sigma, {\slashed{0}}_{15}),  i.e. \Sigb_{4\dot 2^*,4\dot 2^*} =\bar\sigma $, all other component VEVs   zero. If we insist on a cubic tree level  superpotential for $\Phi$  and set also $m=\mu=0$ then in addition to the vanishing singleton VEV in the rank breaking($\Phi_{\nu^c}^{\,\,\nu^c}=\Phi_{\nu^c \nu^{c*}}$) sector  we will have only two possible VEVs emerging from the pair of branch cuts that develop. This case will  the easiest to analyse. Moreover  the quantum  superpotential  $W^{(q)}$  can then also be chosen to be just cubic.

 The symmetry breaking patterns corresponding to the various VEV distribution possibilities can be easily worked out. The easiest way of identifying the unbroken symmetry is to look at the gaugino masses that arise for a given distribution of $v^{(q)}_\pm$ over the 5 VEVs $V_{2-6}$ with $V_1=0, \sigma,\bar \sigma \neq 0$ fixed by the Higgs vacuum structure necessary to reduce the gauge group rank from 5 to 4 without breaking the Standard Model. Firstly it is clear that the nonzero VEVs $\sigma,\bar\sigma$ break $SO(10)\rightarrow SU(5)$ since  16 decomposes as $16=10_1(u^c,e^c,Q_L) + \bar 5_{-3}(d^c,L_L)  +1_5(\nu^c)$ w.r.t. $SU(5)\times U(1)_X,  X=3(B-L) - 4 T_{3R}$.  Thus the VEVs  of $\Sig,\Sigb $ give masses to the 21 gauginos of the coset $SO(10)/SU(5)$ ( $\lambda_{\bar\kappa}^4,\lambda^{\bar\kappa}_4,\lambda_4^4,\lambda_{\dot \alpha \dot\beta}, \lambda_{\bar\kappa 4 \dot 1}^\alpha,\lambda^{\bar\kappa 4 \alpha}_{\dot 2} $  are 22 independent   fields but $\lambda_Y (Y\equiv  2 T_{3R} + B-L)$  remains massless).   The remaining 12  coset gauginos of $SU(5)/G_{321}$   obtain masses unless $V_2=V_3=V_6$ {\emph{and}} $V_4=V_5$. The logic of these conditions is transparent once we note that  $16\times{\overline{16}} = 1 + 45 + 210 $ and 45 and 210 each contain  $SU(5)\times U(1)_X$  singlets corresponding to the  diagonal terms of the  product  $\bar 5(-3) \times 5(3)=1(0)+...$   in the 45-plet getting equal VEVs ($V_4=V_5 $) and the diagonal terms of  product $10(1)\times {\overline{10}}(-1)=1(0) +... $   in the 210-plet getting equal VEVs  $V_2=V_3=V_6$. Thus {\emph{any}} distribution of the quantum VEVs $v^{(q)}_\pm$ over the last 5   ($\Phi_{11}=\Phi_{\nu^c\nu^c*}=0 $ for the Higgs vacuum solution we are focussed on here)  diagonal blocks of $\Phi$ that violates these  equalities will break  $SO(10)\rightarrow G_{123}$. For instance 
 $ \Phi = Diag(0,v^{(q)}_+ {\cal{I}}_4,v^{(q)}_-{\cal{I}}_{11} )$   has $V_2-V_6,V_3-V_6$ both non zero  while   $ \Phi = Diag(0,v^{(q)}_+,  v^{(q)}_-{\cal{I}}_{14} )$  has $V_3-V_6=0$ but $V_2-V_6\neq 0$. Thus {\emph{both}} break to the same SM little group even though the mass patterns are dissimilar.  
  
  The mass spectrum is straightforward to evaluate given the contractions ($A_{\mu\nu \alpha \dot\beta}$ is the  (6,2,2) of $G_{422}$    and  $A^{\mu\nu} _{\alpha\dot\beta} $ is its SU(4) dual)       \cite{alaps}
  \bea 16\cdot 16^* &=&16_{\mu\alpha} (16^*)_{\mu^*\alpha^*}    +16_{\mu^*\dot\alpha} (16^*)_{\mu \dot\alpha^*}   \nnu
  16(\psi)\cdot 16^*(\phi^*) \cdot 45(A) &=& 2 A^\mu_\kappa(\psi_{\mu\beta} \phi^*_{\kappa^*\beta^*} +\psi^\kappa_{\dot\beta} \phi^*_{\mu\dot\beta^*}) -
  {\sqrt{2}} A^{\dot{\alpha}}_{\,\dot\gamma} \psi^\mu_{\dot\alpha} \phi^*_{\mu\dot\gamma^*} \nnu &+& {\sqrt{2}} A^{ \alpha}_{\, \gamma} \psi_{\mu\alpha} \phi^*_{\mu^*\gamma^*}  -A^{\mu\nu \alpha}_{\dot\beta}\psi _{\mu\alpha} \phi^*_{\nu\dot\beta^*} + A_{\mu\nu\beta}^{\dot\alpha}\psi^\mu_{\dot\alpha}\phi^*_{\nu^*\beta^*} \eea
    which can be easily deduced from equations (116)(117) of \cite{alaps} provided we consistently identify
  \bea {\overline{16}}^\mu_\alpha= \epsilon_{\alpha\beta} 16^*_{\mu^*\beta^*} \qquad ;\qquad {\overline{16}}_{\mu \dot\alpha}=- \epsilon_{\dot\alpha\dot\beta} 16^*_{\mu\dot\beta^*} \eea
  The analysis  is straightforward and similar to the   adjoint ($r=N$)  of SU(N) together with   symmetric representations($(N\times N)_s, (\bar N\times \bar N)_s$)   except for the crucial fact that the base representation  16-plet is {\emph{not}} the fundamental of Spin(10).  The evaluation of  the resolvents $R(z),T(z),S(z),U(z)$ and  quantum VEVs proceeds along the lines discussed above  leading again to spontaneous breaking $SO(10) \rightarrow G_{321} $  via dimensional transmutation.  Details will be given in the sequel. We note that the general features of the model are similar to the MSGUT except for the extreme economy with regard to superpotential parameters since the cubic potential for $\Phi$ has just 3 (if $m\neq 0$)   complex parameters and just one in the massless case.  Since  the matter Yukawa couplings $h_{AB},h'_{AB}$  include couplings of pairs of  16-plets to both $10,{\overline{126}}$-plets (symmetric Yukawas) and to $120$-plets(antisymmetric Yukawas) they are general $3\times 3$ complex matrices and thus have ample scope for fitting the observed matter fermion  mass parameters and mixings. Note again the new feature, not present in the MSGUT,   that $\Sig \sim 16\times 16 = 10+ 126 + 120 $ and it is possible to couple the 10, 120 plets generated in this way to the matter 16 bilinear although $16\cdot 16\cdot 126\equiv 0  $ as before.

  \section{Discussion}
  Using  Generalized Konishi Anomaly relations obeyed by gluino and scalar condensates in  supersymmetric vacua  we have shown that   {\emph{asymptotically strong}}  Supersymmetric  Yang Mills Higgs theories  with matrix type Higgs multiplets transforming as general matrix type base $r$ tensors of $G$  provide a calculable implementation of spontaneous symmetry breaking of the   gauge group, including rank reduction,  via dimensional transmutation. This breaking is driven by the formation of    gaugino  condensates in the microscopic/high energy  coupled phase :   which therefore forms an ineluctable background at all larger length scales.  As such they provide a  robust and  novel method of making sense of  AS SuSy GUT models  and justify the surmise that the   UV strong gauge coupling  exhibited by phenomenologically successful and minimal models such as the MSGUT is a signal of nontrivial UV behaviour that makes the theory consistent and yields a sensible low energy limit. This demonstration calls for  deep revision of our notions of the relation between  strong coupling behaviour in the microscopic theory and a phenomenologically acceptable low energy effective theory. 
 
 For any given YM gauge group $G$, the number of asymptotically free models is strictly limited, whereas we have shown that the number of asymptotically strong models with sensible low energy limits is essentially unlimited. Thus  our approach points  the way to a vast expansion of admissible microscopic theories beyond the narrow set of currently canonical    AF type  models. The signal successes of QCD and the amiable ease of analysis of AF models have led, over the half century since their discovery  and  dominance,  to the hardening   of a Dogma that sees AF as the necessary condition for a field theory to be physically  sensible and relevant as a fundamental microscopic  theory.  On the other hand we continue, especially in Condensed matter Physics, to be challenged by the need to tackle quantum systems that are strongly correlated or massively entangled at the microscopic level. The success of the AdS-CFT\cite{adscft} conjecture, and the Seiberg-Witten analysis\cite{seibergwitten} of monopole condensation leading to confinement in $N=2$ supersymmetric YMH theories,   has provided  fruitful working  paradigms  in manifold non-supersymmetric strongly coupled contexts. This  suggests that even our analysis  which is rooted in  supersymmetry  may, in the long run, motivate a  more broad minded view of the way in which microscopic condensation due to strong coupling can generate sensible low energy behaviour.  After all, the strong coupling dynamics underlying satisfaction of the ``kinematic'' GKA constraints must enforce the development of VEVs driven by the physical G-invariant gaugino condensate present at all scales. This phenomenon may well persist even when one moves off the supersymmetric point in coupling space, and even, perhaps, for ``small'' structural differences w.r.t the fields present. These matters require the development of Lattice methods applicable to AS theories for their definitive 
  resolution. The recent development of Lattice methods\cite{LatticeSuSy} applicable to supersymmetric gauge theories encourages us to hope that such methods will be developed. Workers on the lattice will then have a plethora of AS toy models to choose from. For example even the behaviour of our original $SU(2)$ model with a 5-plet of SU(2) (projected out  of   the SU(2) GAM with $r=3$), which is AS,  awaits investigation.
  
  The knowledgeable reader will inquire about the operation of Seiberg duality\cite{seibergduality} in realistic AS models  as also the relation to the ``Asymptotic safety'' program which has attracted much attention lately\cite{litimsann,intrilsann}.  In \cite{berkooz}   Seiberg duals to AF  SO(10) models with several vectors and spinors  were found but extensions to more general representations arising from   tensor products of several vectors/spinors  have remained elusive.  As in Seiberg duality for models in the conformal window\cite{seibergduality}, also  in  the Asymptotic Safety program a non-Gaussian fixed point in the UV  flow of AF of YMH models is required. Recently  the possibility of conformal fixed points in AS SO(10) models such as MSGUT type models was also considered using the so called `a-theorem and c-theorem' constraints  \cite{bajcsann} . However these attempts did not meet any success in finding phenomenologically  viable or   plausible  cases.   We\cite{transuniRG} had earlier shown that RG  fixed points associated with perturbative  gauge   beta function zeros or even Ross-Pendleton type fixed points (in the evolution of ratios of couplings)  do not exist in MSGUT type models because of the huge positive beta functions at one loop.  Even in the absence of non-Gaussian fixed points one may still hope that  AF `magnetic' Seiberg duals to  `electric'  AS MSGUT type models might exist. Such an AF  dual might well exhibit a superconformal fixed point in its IR evolution. However, as mentioned, there has been little progress in finding duals of SO(N) models   with matter in non-fundamental and non-spinor representations so far. 
 
 We have emphasised that in SuSy GUTs, in sharp contrast to say SQCD,  the smallness of the ratio of the SuSy breaking scale $M_S$ to  the GUT scale $M_X$ implies that Supersymmetry may be assumed to be essentially exact at the scales where the theory becomes strongly coupled.  
 Nevertheless it is clear that the issue of supersymmetry breaking must be tackled for such AS GUT models to make contact with reality. The soft supersymmetry breaking terms typically invoked in the MSSM and SuSy GUTs, can be introduced by  spurion Chiral supermultiplets     that take fixed values thus breaking supersymmetry ($S_D=\theta^2 {\bar\theta}^2 m_{\tilde f}^2  , S_M=\theta^2 M_{\tilde g},S_F=\theta^2 A  $  etc. ) and coupling these spurions to Chiral multiplets and Gauge Chiral Field strength appropriately ($ [S_D \Phi^\dagger \Phi]_D, [S_M W_\alpha W^\alpha]_F, [S_A W(\Phi)]_F $ etc.)  yielding SuSy breaking soft terms  for the propagating fields. Since the Konishi anomaly relations involve the lowest components of Chiral multiplets the new terms should not affect the GKA relations directly. After carrying out the GUT SSB  using the quantum VEVs computed by using the CDSW formalism  and thus achieving  robust dimensional transmutation   we  define  a consistent  ``quantum  superpotential ''  $W^{(q)}$ which encodes the quantum VEVs. Using the heavy-light spectrum evaluated therefrom  we can define an effective  low energy supersymmetric model (with exotic operators),  add in the RG run down  small SuSy breaking terms for light fields,  and proceed as usual to study electroweak breaking and low energy phenomenology.  
 
 A related issue is the effect of coupling SYMH to gravity.  Coupling a hidden  sector with SuSy breaking  to the observable sector via gravity, i.e. by considering ${\cal N}=1$ SuperGravity(SuGry)  with YMH,  was the earliest (and still most attractive) method of introducing phenomenologically mandatory  superpartner mass splitting and SuSy breaking trilinear scalar interactions\cite{SuGrysoft}.  Gaugino masses  then arise at two loops from the scalar masses and trilinear couplings even with a standard SYM gauge kinetic function\cite{aulmohnumass}.   On the other hand, in  a {\emph{fermionic}} background  provided by a  non-vanishing Gravitino Field strength (the lowest member of the so called ${\cal N}=1$  Weyl multiplet $G_{\alpha\beta\gamma}$ ) the Chiral ring  is modified and the factorization of correlators receives\cite{gravchiring} corrections of $O(G^2)$.  However such backgrounds are hardly of any phenomenological interest from the point of view of GUTs   since they can presumably be significant   only in regions of  Planck  density.  On the other hand, with the quantum  corrected superpotential $W^{(q)}$  we may proceed in the standard way by embedding in ${\cal N}=1$ SuGry and deriving soft SuSy breaking terms as usual  by adding say a Polonyi SuSy breaking superpotential using a singlet scalar field. No modification of the GKA analysis is called for in such a hybrid approach. 
 
One novel and mysterious implication of the GKA relations is that even the superpotential terms containing matter chiral multiplets(along with Higgs multiplets)  must participate at least in trilinear condensates with superheavy values, even though VEVs for the smatter fields are phenomenologically unacceptable. The phenomenological    implications  of such three point correlators  are not clear to us.  Perhaps such novel quantum background contaminations of the perturbative theory  will eventually yield novel signals of the dynamical symmetry breaking origin of GUT spontaneous symmetry breaking.

 \vspace{1 true cm}

{\bf{Acknowledgment :}}
It is a pleasure to thank   P.Ramadevi,  T.Enkhbat, K.S. Narain, A. Joseph, K.P. Yogendran  and  especially  B. Bajc  for discussions,  and  P. Guptasarma for encouragement. The support of the ICTP, Trieste Senior Associates program through  an award during 2013-2019  and the  hospitality of the ICTP High Energy Group, and in particular G. Senjanovic,   both when this idea was first conceived in 2002 and during the summer of 2019, is gratefully acknowledged.

\end{document}